\documentstyle[epsfig]{mn}
\newcommand{\be}{\begin{equation}}
\newcommand{\ee}{\end{equation}}
\newcommand{\bea}{\begin{eqnarray}}
\newcommand{\eea}{\end{eqnarray}}
\newcommand{\bc}{\begin{center}}
\newcommand{\ec}{\end{center}}

\newcommand{\lu}{\,h^{-1}{\rm Mpc}}

\title{Recovering the Topology of the Initial Density Fluctuations Using the {\em IRAS} Point Source Catalogue Redshift Survey}

\author[A. Canavezes J. Sharpe]{A. Canavezes,$^1$ J. Sharpe,$^2$ \\
$^1$ Centro de Astrof\'{i}sica da Universidade do Porto, Rua das Estrelas s/n, 4150 Porto, Portugal\\
$^2$ Imperial College of Science Technology and Medicine, Blackett Laboratory, Prince Consort Road, London SW7 2BZ, UK}

\begin{document}
\maketitle

\begin{abstract}

We apply the reconstruction technique of Nusser \& Dekel (1992) to the recently available Point Source Catalogue Redshift Survey (PSCz) in order to subtract the phase correlations that are expected to develop in the mild non-linear regime of gravitational evolution. We study the evolution of isodensity contours defined using an adaptive smoothing algorithm in order to minimize the problems derived from the non-comutivity of operators. We study the genus curves of these isodensity contours and concentrate on the evolution of the amplitude drops, a meta-statistic able to quatify the level of phase-correlation present in the field. In order to test the method and to quatify the level of statistical uncertainty, we apply the method to a set of mock PSCz catalogues derived from the N-body simulations of two 'standard' CDM models, kindly granted to us by the Virgo consortium. We find the method to be reliable in recovering the right amplitude drops. When applied to PSCz the level of phase correlations observed is very low on all scales ranging from $5\lu$ to $60\lu$, providing support to the theory that structure originated from gaussian initial conditions.

\end{abstract}
 
\begin{keywords}
galaxies:clusters:general -- cosmology:observations -- cosmology:large-scale-structure of the Universe
\end{keywords}

\section*{INTRODUCTION}

        \label{rec}

\begin{figure*}
\bc
\resizebox{14cm}{!}{\includegraphics{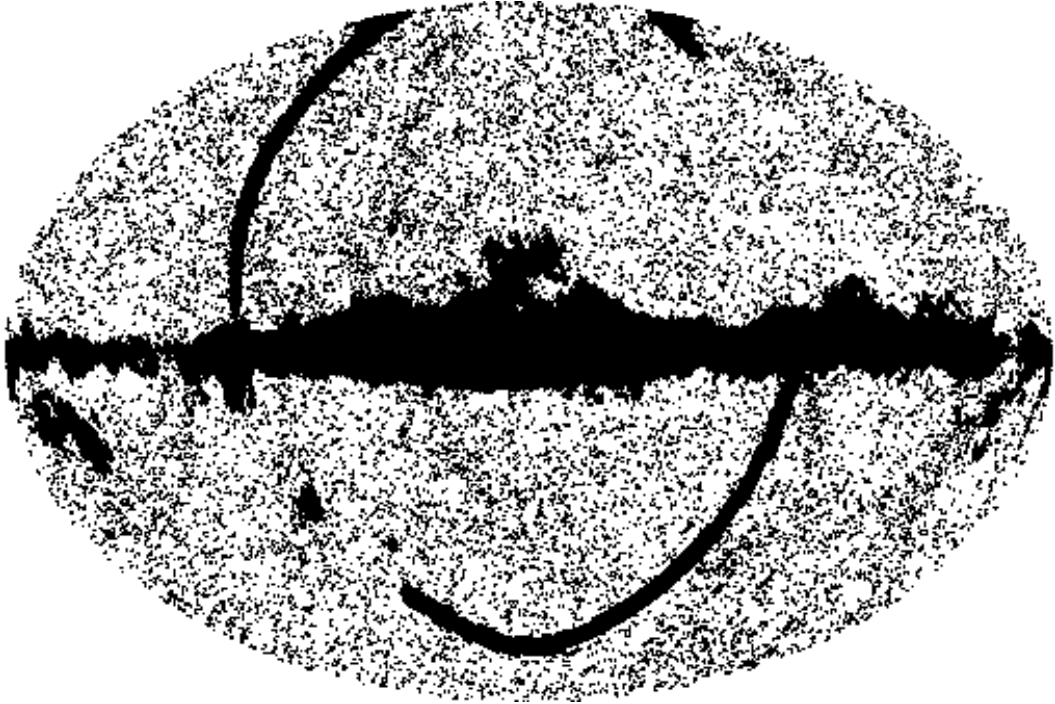}}
\caption
{Sky coverage of the PSCz survey in an aitoff projection. The dots represent the galaxies in the survey, the shaded regions are unobserved and comprise the angular mask. The galactic centre lies in the centre of the plot.
\label{figMask}}
\ec
\end{figure*}

The {\em IRAS} Point Source Catalogue Redshift Survey \cite{Sa99} lends itself to topological studies because of its high sampling densities and large volume. It countains approximately $15000$ galaxies to the full depth of the PSC ($0.6 Jy$). Its sky coverage is $84.1$ per cent, where only the zone of avoidance is excluded, here defined as an infrared background exceeding $25 \rm{MJy sr}^{-1}$ at $100\mu\rm{m}$, and a few unobserved or contaminated patches at higher latitude. The excluded regions are coded in an angular mask, as shown in Fig. \ref{figMask}. Canavezes et al. \shortcite{Ca98} analysed the topology of PSCz and showed that it is consistent with the topology of CDM models: The genus curves retain the w-shape characteristic of random-phase density fields even at small smoothing lengths, where non-linear evolution has already generated significant skewness of the 1-point PDF. These non-linearities are, however, detected by a depressed amplitude of the genus curves - amplitude drops - consistent with those detected for the CDM models. Above $10 \lu$ strong phase correlations were not detected in PSCz and those that were detected at smaller scales are expected in the framework of mildly non-linear gravitational evolution. This supports the hypothesis that structure grew from random-phase initial conditions.

In this paper we attempt to subtract these phase correlations by reversing gravity. If it is true that structure does indeed originate in random-phase fluctuations, then reversing gravitational evolution should provide us with an initial density fluctuation field free from phase correlations on any scale.

There are several methods to achieve this goal. Nusser \& Dekel \shortcite{ND92}, showed how to express the Zel'dovich approximation in a set of Eulerian coordinates and reverse it to obtain a reconstructed density field. They tested the method with N-body simulations and found satisfactory results. Recently, some improvements have been made on the original method (e.g. Gramann \shortcite{Gr93}). Nusser, Dekel \& Yahil \shortcite{Nu95} applied the Nusser \& Dekel \shortcite{ND92} approximation to the $1.2$ Jy $IRAS$ redshift survey to recover the 1-point probability distribution function of the initial density field. Their results were consistent with Gaussian initial conditions. The PSCz presents us with an unprecedented number of resolution elements and should constitute therefore the best available data set on which to apply a reconstruction technique. 

The method we choose to follow here is the quasi-linear method proposed by Nusser \& Dekel \shortcite{ND92}. In section \ref{recmethod} we describe this method; In section \ref{recdenmaps} we describe the construction of the density maps used as the input of the {\em time-machine} and describe our error estimates; Section \ref{recnbody} is devoted to testing the method using N-body simulations of  a CDM model and in section \ref{recpscz}  we apply the method to the PSCz and present our results.

\section{THE RECONSTRUCTION METHOD}
\label{recmethod}

The reconstruction method we follow is based on two premises: The velocity field is, when smoothed on a scale of a few Mpc, irrotational, and the Zel'dovich approximation is accurate over the mildly non-linear regime. 

Our first step is to try to express the Zel'dovich approximation in a set of Eulerian coordinates.

Given an initial comoving position $\bf{q}$ for a given particle, a final position $\bf{x}(\bf{q},t)$ will have the form
\be 
{\bf x}({\bf q},t)={\bf q}+P({\bf q},t).
\ee
The Zel'dovich approximation states that the displacement term $P({\bf q},t)$ can be written as a product of two functions, each a function only of one of the variables ${\bf q}$ or $t$, i.e., we can separate the variables $q$ and $t$:
\be
{\bf x}({\bf q},t)={\bf q}+D(t)\Psi({\bf q}).
\ee
This is simply a linear approximation with respect to the particle displacements rather than density. Particles in the Zel'dovich approximation follow straight lines:
\be
{\bf v}({\bf q},t)=a(t)\frac{d{\bf x}}{dt}=a(t)\dot{D}(t)\Psi({\bf q}).
\ee
The density fluctuation is given by
\be
\delta({\bf q},t)=\bar{\rho}/J({\bf q},t)-1,
\ee
where $J({\bf q},t)$ is the Jacobian of the coordinate transformation ${\bf q}\rightarrow {\bf x}$.

As long as second order terms in $\delta$ and ${\bf v}$ are negligible, $\delta(t) \propto D(t)$, where $D(t)$ is the growing mode solution in linear perturbation theory (see e.g. Peebles \shortcite{Pe80}).

Unlike linear theory, where the density at a given position evolves according to the linear growth rate, under the Zel'dovich approximation infinite density can develop in a finite time as a result of the convergence of particle trajectories into a {\em pancake}. in order to integrate a density field back in time, Nusser \& Dekel \shortcite{ND92} found the differential equation in Eulerian space which contains the Zel'dovich approximation. This is derived from the standard equation of motion of dust particles in an expanding Universe (e.g. Peebles \shortcite{Pe80}):
\be
\frac{d{\bf v}}{dt} + H{\bf v}=-\frac{1}{a}\nabla\Phi_{g} .
\label{eqmotion}
\ee
Here ${\bf v}$ stands for the {\em peculiar} velocity of the particle, i.e., ${\bf v}=a(t)\frac{d{\bf x}}{dt}$, where ${\bf x}$ are the comoving positions; $\nabla=\left( \partial /\partial x,\partial/\partial y,\partial /\partial z \right)$  and $\Phi_{g}$ is the gravitational potential which is related to the local density fluctuation via the Poisson equation
\be
\nabla^{2}\Phi_{g}=\frac{3}{2}H^{2}\Omega a^{2} \delta .
\label{poisson}
\ee

Using the normalized variables $\theta$ and $\varphi_{g}$ defined thus:
\begin{eqnarray}
\theta({\bf x},t)&\equiv &\frac{{\bf v}({\bf x},t)}{a\dot{D}}={\bf \psi}({\bf q})  \label{renormtheta}
\\
\phi_{g}({\bf x},t)&\equiv & \frac{\Phi_{g}({\bf x},t)}{a^{2}\dot{D}} \label{renormphi} ,
\end{eqnarray}
equations \ref{eqmotion} and \ref{poisson} reduce to
\be
\dot{\theta} + \frac{3H\Omega}{2f(\Omega)}\theta=-\nabla\varphi_{g} ,
\label{motiontheta}
\ee
where $f(\Omega)\equiv \dot{D}/HD \sim \Omega^{0.6}$ \cite{Pe80}.

Notice that the Zel'dovich approximation is contained in \ref{renormtheta}, which is to say 
\be
\dot{\theta}=0
\label{zeld}
\ee
 along the trajectory of a given particle, i.e., along a line of constant ${\bf q}$. Hence, \ref{motiontheta} is reduced, under the Zel'dovich approximation, to
\be
\frac{3H\Omega}{2f(\Omega)}\theta=-\nabla\varphi_{g} .
\ee
$\theta$ is therefore the gradient of a potential:
\be
\theta=-\nabla\varphi_{v},
\label{gradient}
\ee
with
\be 
\phi_{v}=\frac{2f(\Omega)}{3H\Omega}\varphi_{g} + F(t) .
\ee
We can obviously set $F(t)\equiv 0$ since $\theta$ is the only physical (measurable) quantity.

Equation \ref{zeld} can now be expanded in Eulerian coordinates:
\begin{eqnarray}
\frac{d\theta}{dt}=0&\Leftrightarrow & \left[ \frac{\partial}{\partial t} + \frac{d{\bf x}}{dt}\cdot \nabla \right] \theta({\bf x},t) = 0 \nonumber \\
& \Leftrightarrow & \frac{\partial \theta}{\partial t} + \dot{D}(t) {\bf \psi}({\bf q}) \cdot \nabla \theta=0 \nonumber \\
& \Leftrightarrow & \frac{\partial \theta}{\partial t} + \dot{D} \left( \theta\cdot \nabla \right) \theta =0 .
\end{eqnarray}
Making use of the identity $\left( \theta\cdot \nabla \right) \theta = 1/2 \left( \nabla \theta \right) ^{2} - \theta \wedge \left( \nabla \wedge \theta \right) $, taking into account the irrotationality of $\theta$, and substituting \ref{gradient} we arrive at
\begin{eqnarray}
-\nabla \frac{\partial \varphi_{v}}{\partial t} + \dot{D}\frac{1}{2} \left(\nabla \theta\right)^{2} =  0 & \Leftrightarrow & -\nabla \left[ \frac{\partial \varphi_{v}}{\partial t} + \frac{\dot{D}}{2} \left(\nabla \varphi\right)^{2} \right] =  0 \nonumber \\
& \Leftrightarrow &  \frac{\partial \varphi_{v}}{\partial t} + \frac{\dot{D}}{2} \left( \nabla \varphi_{v}\right) ^{2}  = F(t) .
\end{eqnarray}
Again, because only gradients of $\varphi_{v}$ have a physical meaning, we can take $F(t)$ to be zero. We finally arrive at the equation:
\be
\frac{\partial \varphi_{v}}{\partial D} + \frac{1}{2} \left( \nabla \varphi_{v}\right) ^{2} =0 .
\label{motion}
\ee 

This is the differential equation which expresses the Zel'dovich approximation in Eulerian coordinates. Knowing $\varphi_{v}$ at any time enables us to compute $\varphi_{v}$ at any other time simply by integrating \ref{motion} forwards or backwards.

The first step is to calculate $\varphi_{v}$ at the present time. That can be easily achieved using Poisson's equation, which in Fourier space reads:
\be
-k^{2} \tilde{\Phi}_{g}=\frac{3}{2}H^{2}\Omega a^{2} \tilde{\delta} ,
\ee
where the tildes denote the Fourier transforms. Using \ref{renormphi} we arrive at:
\be
-Dk^{2}\tilde{\varphi}_{v}=\tilde{\delta} .
\label{potenti}
\ee
So, in order to obtain the velocity potential of a given smooth density field with periodic boundary conditions, we first transform it to Fourier space using some FFT code, we then divide the obtained field by $-k^{2}$ (normalizing D to unity at present), and then transform back to real space to obtain $\varphi_{v}$. $\varphi_{v}$ is then integrated back in time in the most trivial way: At every step, we calculate $1/2 \left( \nabla \varphi_{v} \right) ^{2}$ from the potential field; this is then used as a first order Taylor correction to predict the value of $\varphi_{v}$ at the next step. Once a suitable number of integration steps has been performed, we use equation \ref{potenti} again to obtain the value of $\delta$ at the corresponding value of $D$.

For our application - topology, we expect the exact number of integrations
to be irrelevant.  After a certain number of iterations we expect to reach the linear
regime in all scales, which means that the genus curves of the density fields will no longer change. This particular characteristic of the genus can in fact be useful to test the convergence of the method. We obtained several reconstructed density fields for different values of the number of integration steps, and found the topologies to converge. This result is shown is Fig. \ref{convergence}.

\section{CONSTRUCTION OF THE DENSITY MAPS}
\label{recdenmaps}

The construction of the density maps we follow is, to a large extent, equivalent to the method followed by Canavezes et al. \shortcite{Ca98}. We employ the same fit for the PSCz selection function $s(z)$:
\be
s(z)=\frac{\psi}{z^{\alpha}\left( 1+\left( \frac{z}{z^{\star}}\right) ^{\gamma} \right)^{\beta/\gamma}} ,
\ee
with the parameters shown in Table \ref{tab1} \cite{Sp98a}, and estimate the density $\rho(r)$ by
\be
\rho(r)\propto\frac{m(r)}{s(r)} ,
\ee

\begin{table}
\bc
\caption{Parameters of the selection function of PSCz.\label{tab1}}
\vspace{0.5cm}

\begin{tabular}{ccc}
$\alpha$  & $\beta$ & $\gamma$ \\
$0.991^{+0.068}_{-0.073}$  
&  $3.445^{+0.173}_{-0.158}$ 
& $1.925^{+0.162}_{-0.153}$ \\
\\
$z^{\star}$ & $\psi \;\;[h^3{\rm Mpc}^{-3}]$ & \\
$0.02534^{+0.00130}_{-0.00116}$ 
&$(141.3\pm 2.4 ) \times 10^{-6}$ & \\
\end{tabular}
\ec
\end{table}

\begin{table}
\bc
\caption{The smoothing lengths adopted for the topological analysis of the PSCz
survey. Listed are the adopted survey depth $R_{\rm max}$, the
resulting number
$N_{\rm res}$ of resolution elements and the number $N_{\rm gal}$ 
of galaxies inside the survey volume.
\label{tab2}
}
\vspace{0.5cm}
\begin{tabular}{c|c|r|r|r}
$\lambda\;\;[\lu]$  & $R_{\rm max}\;\;[\lu]$  &
$N_{\rm res}$ & $N_{\rm gal}$  \\
 5&  34.92&  215.5&    2295\\
 6&  47.16&  307.3&    3550\\
 7&  58.35&  366.3&    4928\\
 8&  68.58&  398.5&    5909\\
10&  86.91&  415.3&    7510\\
12& 103.19&  402.3&    8775\\
14& 118.04&  379.2&    9681\\
16& 131.81&  353.7&   10356\\
20& 156.96&  305.8&   11309\\
24& 179.79&  266.0&   11968\\
28& 200.97&  233.9&   12496\\
32& 221.79&  210.6&   12833\\
40& 263.44&  180.7&   13339\\
48& 305.09&  162.4&   13684\\
50& 315.50&  158.9&   13748\\
56& 346.73&  150.2&   13924\\
\end{tabular}
\ec
\end{table}

where $m(r)$ is the discrete point distribution. However, the angular mask needs to be treated with care. Because the structure we can see behind the mask will have a significant effect on the evolution of structure on the whole observable area, we need to resort to some sort of filling, prior to using the time-machine. 

We consider two different types of filling: One, a random filling, in which fake objects are placed randomly over the masked region with an average number density equal to the average number density of the whole survey (weighted by the selection function); And another, a {\em cloning}, in which fake objects are placed randomly in each bin, but with an average number density equal to the average number density of the neighbouring observable bins, again weighted by the selection function. 

On both cases, we create a box containing a sphere of radius $R_{max}$ as defined in Table \ref{tab2}. $R_{max}$ is the maximum distance up to which the average distance between two neighbouring galaxies in PSCz is smaller than the adopted smoothing length. The size of the box will then be $\frac{2}{\sqrt{3}}Rmax$. Although boundary conditions are not periodic, we smooth the density field in Fourier space assuming periodic boundary conditions. We consider this to be preferable to zero padding, as this would create an artificial boundary that would eventually affect the gravitational time-machine.

We then apply the Zel'dovich time-machine on the box thus obtained for both filling techniques. In the subsequent topological analysis, we limit ourselves to the sphere inside the box with radius $R_{max}$. 

\subsection*{Smoothing Procedure}

Ideally, one would want to apply the operator {\em time-reverse} directly upon the unsmoothed density field obtained from PSCz, and only then smooth it on a range of scales to be able to calculate its genus curve. However, in order to obtain meaningful results, we need a smooth field {\em a priori}. This poses a very important problem: These operators ({\em time-reverse} and {\em smoothing}) do not in general commute. In other words, the topology of the final density field can be significantly different whether the smoothing operation is performed before or after applying the {\em time-machine} to our original field.

To illustrate this point, let us consider a Gaussian random density field smoothed on some scale $\lambda$. After applying the {\em time-machine} this density field will still be Gaussian. However, if we are to naively calculate its genus curve choosing for isodensity contours the same isodensity contours defined {\em prior} to applying te {\em time-machine}, we would wrongly conclude that the field is not Gaussian. In order to circumvent this problem we would be forced to smooth the final field once again on a larger scale, thus reducing considerably the statistical significance of our results.

There is, however, an alternative way to solve this dilemma: One can try to find a smoothing operator that comutes with our reconstruction operator.

How can we look for such a smoothing operator? One way of ensuring this is to find an operator that does not change  the density field if applied once again at a latter stage. By other words, if we ensure that our final density field remains the same regardless of how many times we perform the smoothing operation throughout the time-reversing process, then it is because these two operators do indeed comute. A class of such operators are the adaptive smoothing operators. By adaptive smoothing, we understand a {\em local} smoothing of variable smoothing length according to the local structure. There are several types of adaptive smoothing. The ideal adaptive smoothing operator ensures that the total mass enclosed by isodensity contours remains constant throughout the time-reversing process. A simplified version assumes spherical symmetry. At each point we employ a smoothing length $\lambda$ such that $\lambda^{3}\propto 1/\rho=1/\left( \rho_{0}+\delta \right)$. The proportionality constant defines the characteristic smoothing length, which is the smoothing length employed when the overdensity $\delta$ vanishes. For highly dense regions, a smaller value of $\lambda$ will be chosen, whereas voids will be smoothed with a larger value of $\lambda$.

When we apply the time-machine operator on a map smoothed in this way, the isodensity contours around a cluster will move, but the total mass enclosed by them will remain approximately constant, as long as the spherical symmetry hypothesis is a good approximation. 

In our application to the PSCz we use a spherically symmetric adaptive smoothing algorithm. We start by creating a set of maps obtained from the original map (this might be either the PSCz data or simulation data) by smoothing it on a range of scales around a given characteristic scale $l_{0}$, using Gaussian filters of the form
\be
G_{\lambda}(x)=\frac{1}{\pi^{3/2}\lambda^{3}} e^{-x^{2}/\lambda^{2}}
\ee
where $\lambda$ varies around $l_{0}$. We then look for the appropriate value of $\lambda$ at each position $x$ by enforcing the equation
\be
(1+\delta_{\lambda})\lambda^{3}=l_{0}^{3} ,
\ee  
where $\delta_{\lambda}$ is the density contrast when the original map is smoothed on a scale $\lambda$. This ensures that the mass enclosed on a sphere of radius $\lambda$ is, to first order in a Taylor expansion, constant throughout the whole map.

\subsection*{Error Estimates}

There are three different sources of error that enter our results: Shot noise, cosmic variance and the errors associated with the Zel'dovich approximation itself.

The most accurate and realistic way of estimating these errors, which ultimately enter the genus curves of the reconstructed PSCz density fields, is to analyse the topology of fake PSCz catalogues drawn from N-body simulations of some {\em standard} CDM model. Since we intend to test the validity of the Zel'dovich time-machine with N-body simulations, it seems reasonable to extent this philosophy to the calculation of the statistical errors themselves. This is achieved by seeking a galaxy number density field with a Poisson distribution, whose expectation value is identical to the density field of the N-body simulation multiplied by the PSCz selection function. This is equivalent to {\em observing} the density field of the N-body simulation in a similar way to the way PSCz observes the density field of the real Universe. This galaxy number density field is then divided by the PSCz selection function again to obtain a "PSCZ-noisy" distance-independent estimate of the real galaxy number density. Each of the {\em mock} PSCz catalogues is adaptively smoothed using the algorithm described above and then used as input in the Zel'dovich time-machine. The genus curves of the reconstructed fields are calculated and the variance obtained over 10 mock PSCz catalogues is used as the statistical error estimate.

\section{TESTING THE METHOD WITH N-BODY SIMULATIONS OF CDM MODELS.}
\label{recnbody}

As we mentioned previously, we intend to test the regime of validity of the Zel'dovich time-machine by means of an N-body simulation of a CDM model. We apply the time-machine on the present density field drawn from the simulation and compare the density field thus obtained with the density field drawn from the original test-mass positions used as input in the simulation. More specifically, we need only to compare the topologies of the reconstructed and original fields  through their genus curves.

For this purpose, we use two N-body simulations corresponding to two cold dark matter models, kindly provided by the Virgo consortium \cite{Col97,Je97}.The simulations have been performed with an  AP$^{3}$M-SPH code 
named {\small HYDRA} \cite{Cou95}. Here we consider the SCDM model and the $\tau$CDM model, which parameters are shown in Table \ref{modelparameters}. Because this simulations contain CDM on periodic boxes of size $239.5\lu$ and use such a large number of particles, they constitute an ideal ground for this test.

\begin{table}
\bc
\caption{Parameters of the examined CDM models. The simulations have
been done by the 
Virgo collaboration.\label{modelparameters}
}
\begin{tabular}{l|c|c|c|c|}
\multicolumn{1}{l|}{ }& $\tau$CDM & $\Lambda$CDM
\vspace{0.1cm}\\ 
\multicolumn{1}{l|}{Number of particles } & $256^{3}$ & $256
^{3}$ \\
\multicolumn{1}{l|}{Box size$[\lu]$ } & $239.5$ & $239.5$\\
$z_{start}$& $50$ & $30$ & \\
$\Omega_{0}$ & $1.0$ & $0.3$ \\
$\Omega_{\Lambda}$& $0.0$ & $0.7$\\
 \multicolumn{1}{l|}{Hubble constant $h$} & $0.5$ & $0.7$\\
$\Gamma$ & 0.21 & 0.21 \\
$\sigma_{8}$ & $0.60$ & $0.90$ \\
 \multicolumn{1}{l|}{Mass per particle [$10^{10}h^{-1}M_{\odot}$]} & $22.7$ &$5.8$ & \\

\end{tabular}
\ec
\end{table}

We start by binning the particles in cells of side length $2\lu$ and smooth further on some characteristic smoothing length $\lambda$, using the adaptive smoothing algorithm described in section \ref{recdenmaps}, in an analogous procedure to the way we treat the PSCz galaxies. This "double" smoothing is required in the real Universe in order to minimize shot noise effects, but it is also necessary in order to eliminate severe non-linearities since the Zel'dovich time-machine is expected not to work over such regimes. It is also needed to smooth out regions of orbit-crossing.

The next step is to calculate the genus curves of both the original density field that is used as input in the simulation, smoothed in the same manner as described above, and the reconstructed density field obtained after applying the Zel'dovich time-machine to the present density field smoothed in the same way. In order to compare both genus curves we compute their amplitudes and their amplitude drops. It is very important to note that if we are to obtain gaussianized versions of the density fields at the present time, they will not have, in principle, the same genus amplitudes as the density fields themselves, even when these are in the Gaussian regime. This is so because of the very nature of adaptive smoothing. When a given density field is smoothed adaptively its genus curve will have a different shape and amplitude than the shape and amplitude of the genus curve of the same field when this is smoothed with a constant Gaussian window. We need to be extremely careful not to draw the wrong conclusions about the Gaussian or non-Gaussian nature of our density fields. After applying the time-machine, however, the amplitude of fluctuations will be reduced significantly. In fact they will be reduced to such an extent as to make the field look almost homogeneous. This means that the isodensity contours on adaptively smoothed maps will be very close to the isodensity contours on maps smoothed using a constant Gaussian window. Hence, it is possible to calculate amplitude drops and determine the Gaussian (or non-Gaussian) nature of our density maps {\em after} applying the Zel'dovich time-machine, i.e., to determine the Gaussian (or non-Gaussian) nature of the {\em reconstructed} fields.

\begin{figure*}
\bc
\resizebox{6cm}{!}{\includegraphics{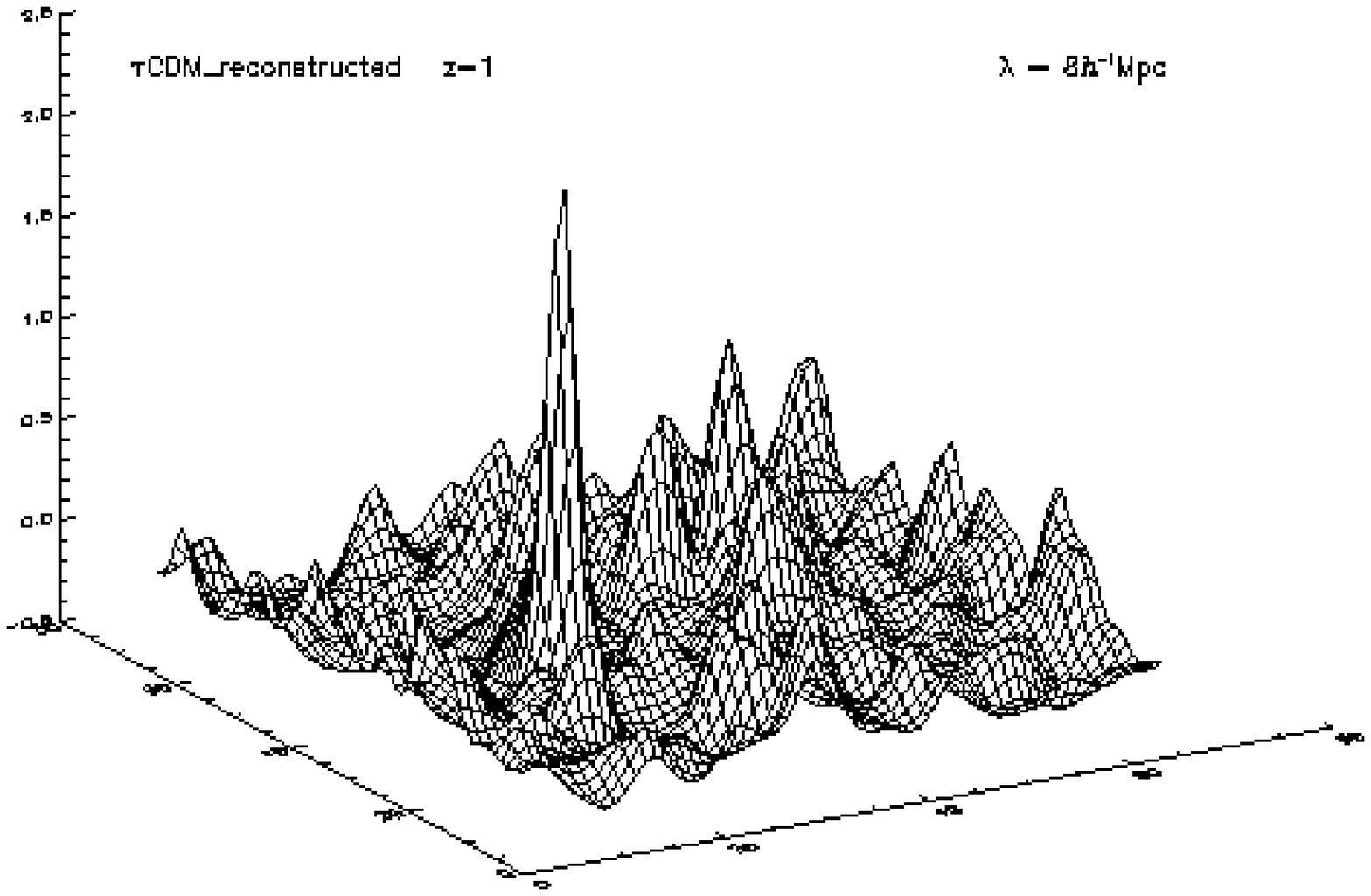}}
\resizebox{6cm}{!}{\includegraphics{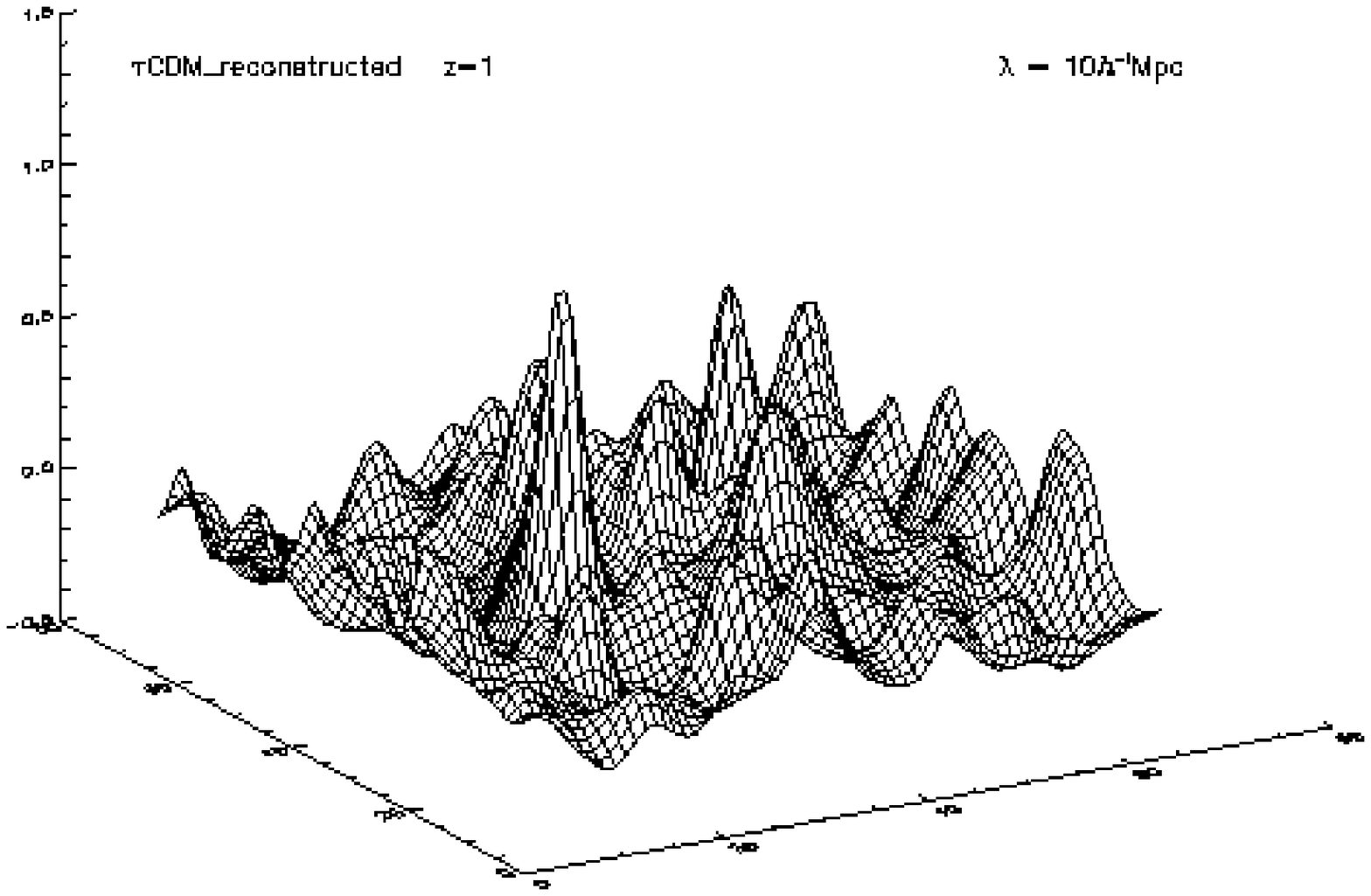}}
\resizebox{6cm}{!}{\includegraphics{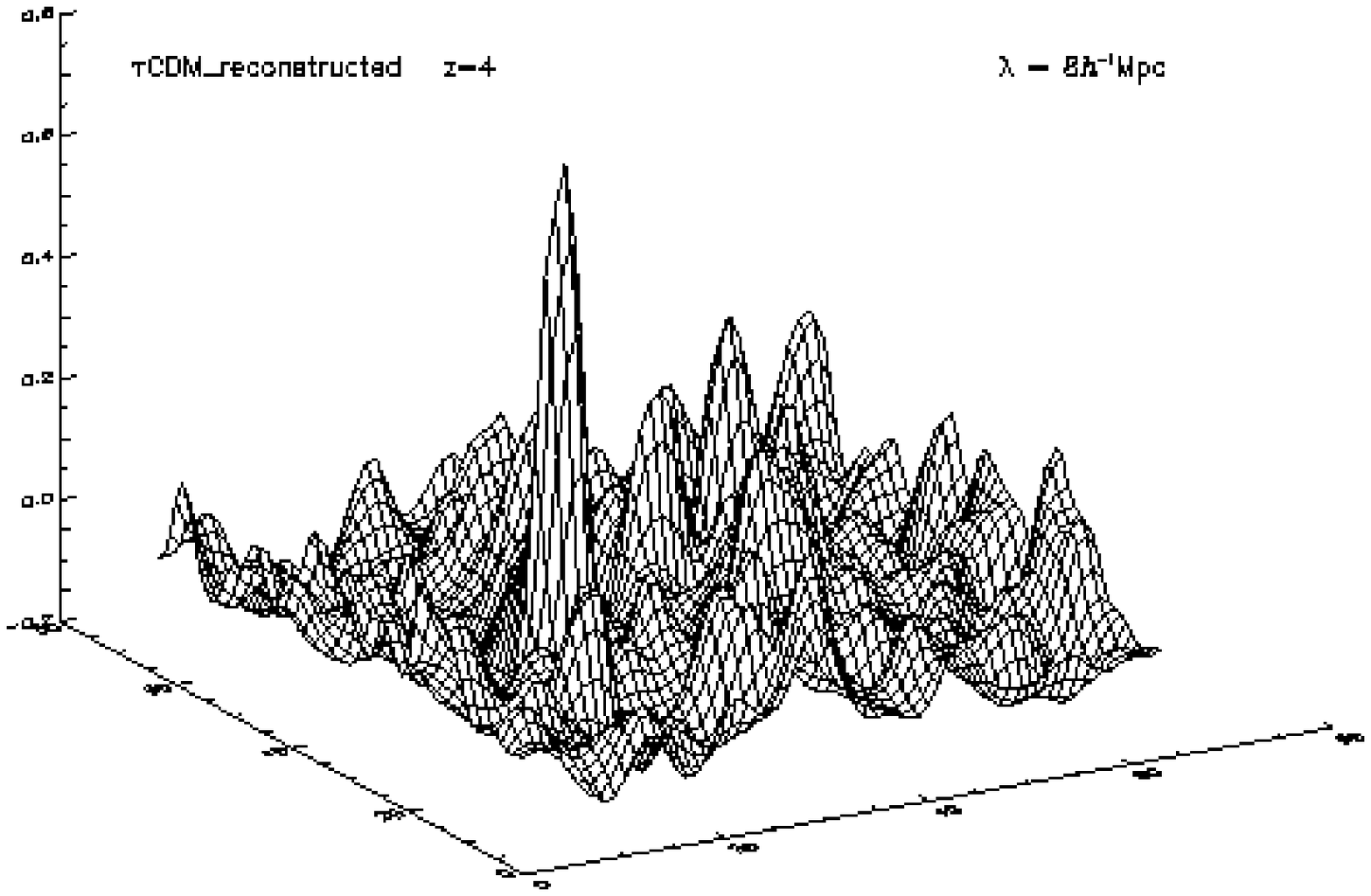}}
\resizebox{6cm}{!}{\includegraphics{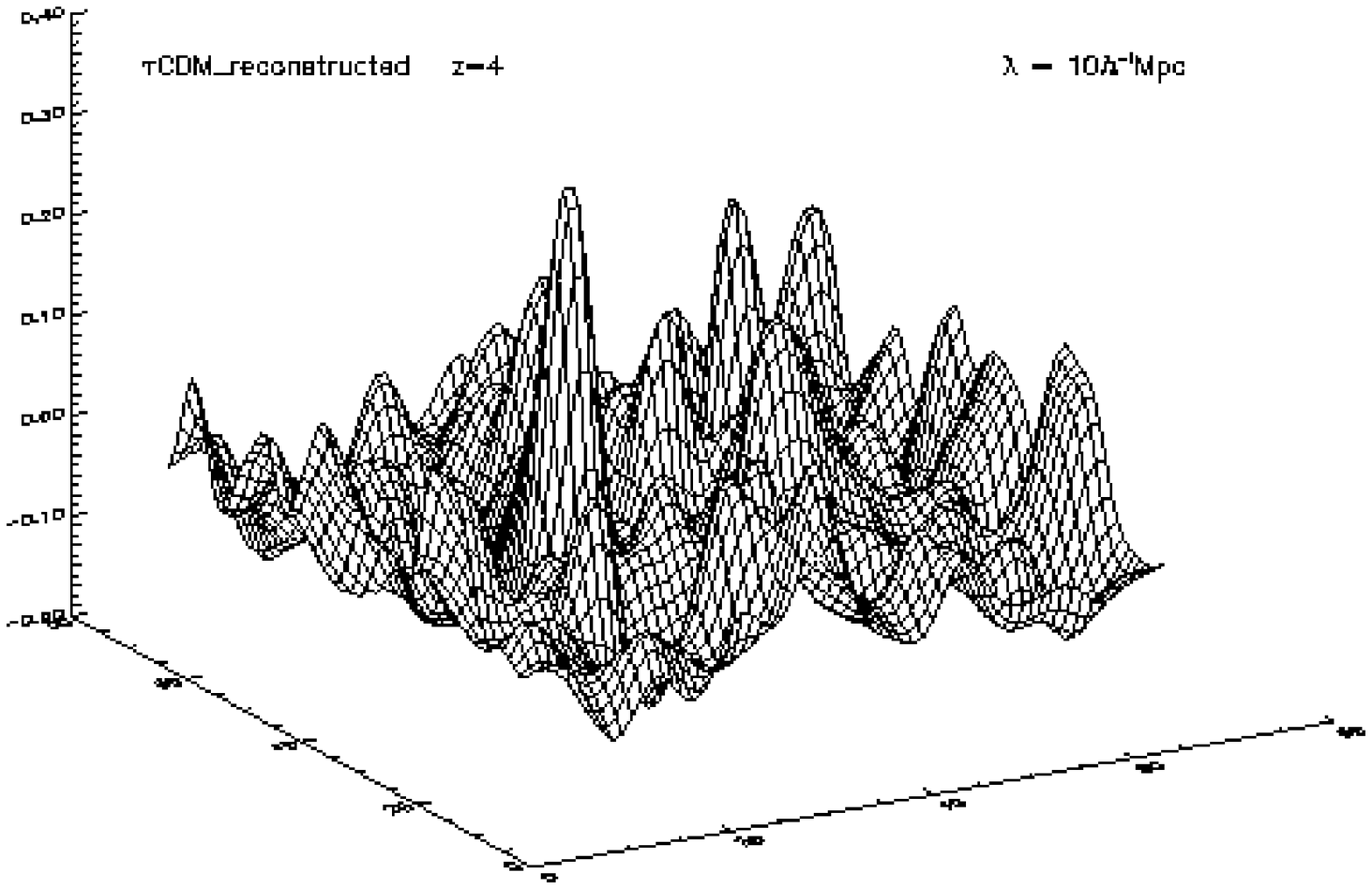}}
\resizebox{6cm}{!}{\includegraphics{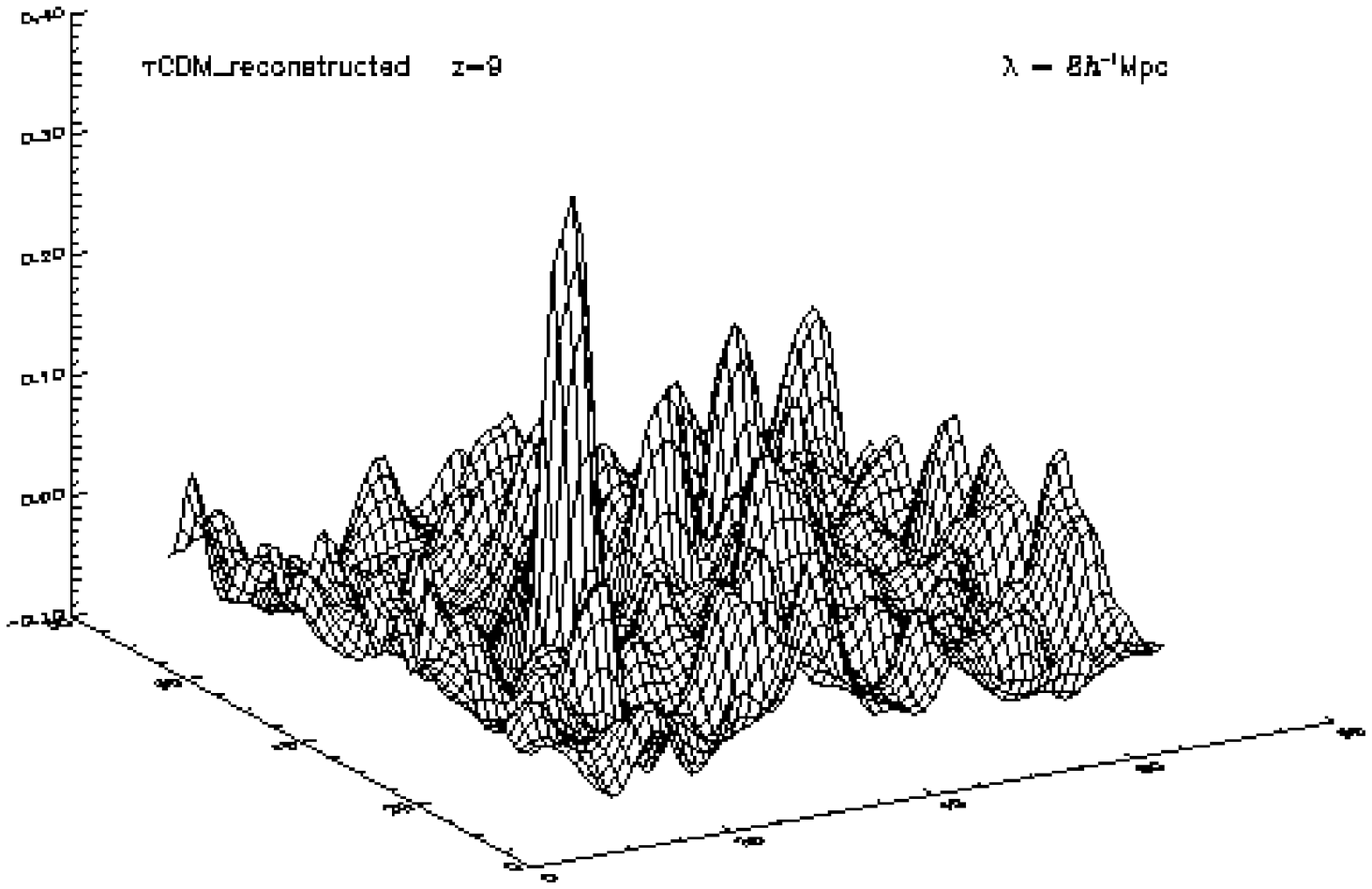}}
\resizebox{6cm}{!}{\includegraphics{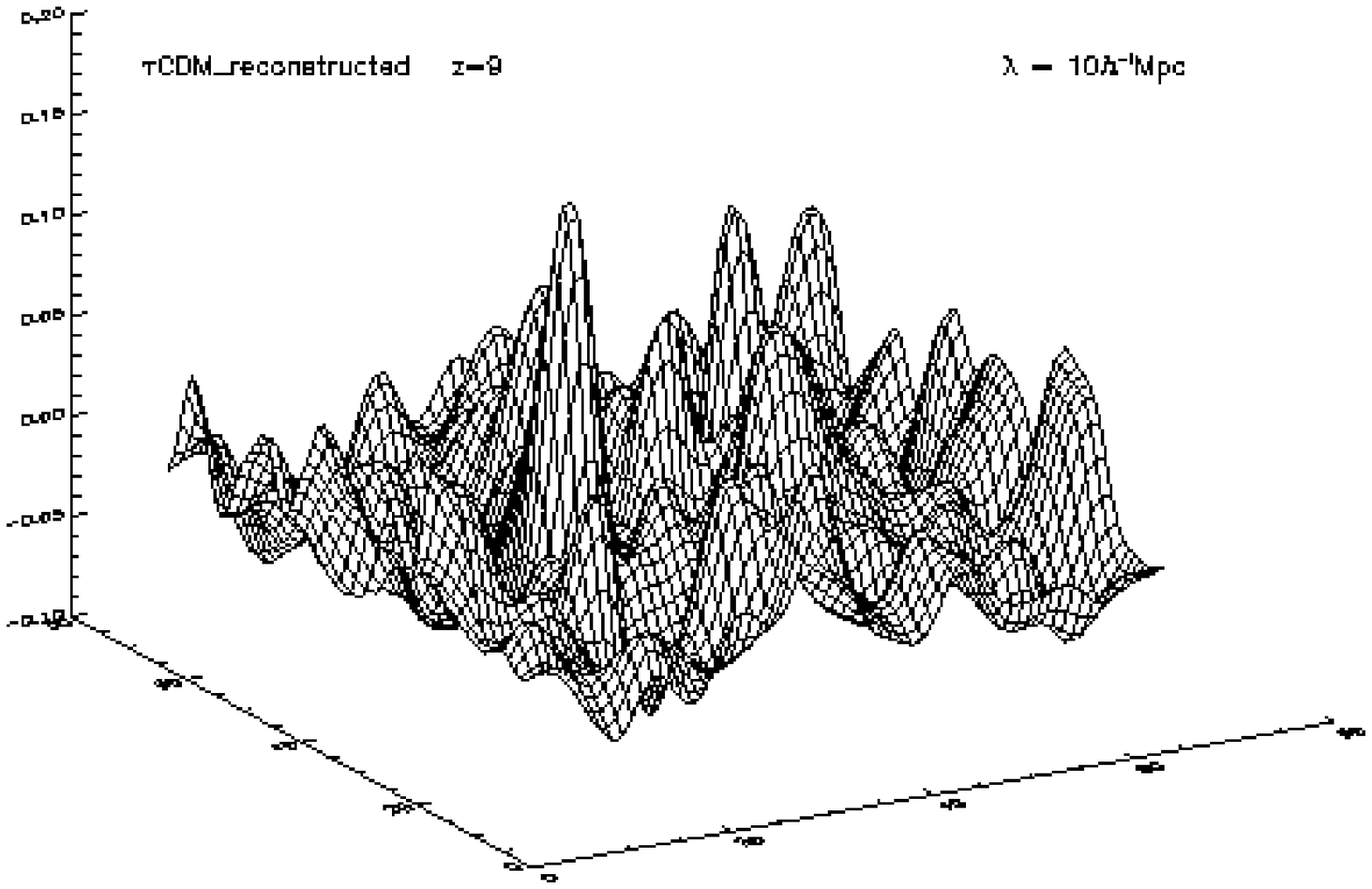}}
\caption{A particular slice of the reconstructed fields for the $\tau$CDM simulation for different values of $z$. The left panels show the fields smoothed adaptively with a characteristic smoothing length of $8\lu$, whereas the panels on the right show the fields smoothed with a characteristic smoothing length of $10\lu$ \label{convergence}}
\ec
\end{figure*}

\begin{figure*}
\bc
\resizebox{8cm}{!}{\includegraphics{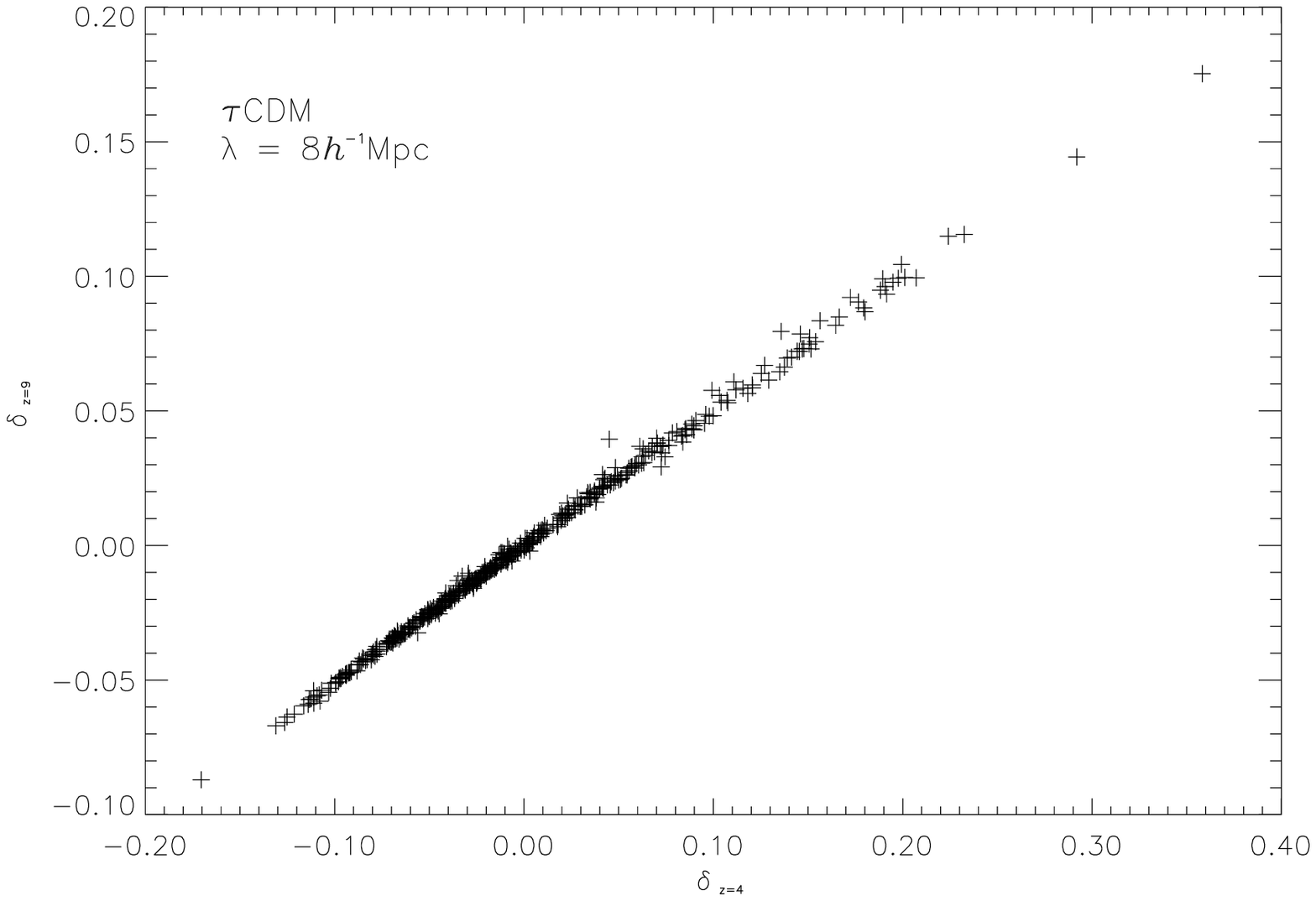}}
\resizebox{8cm}{!}{\includegraphics{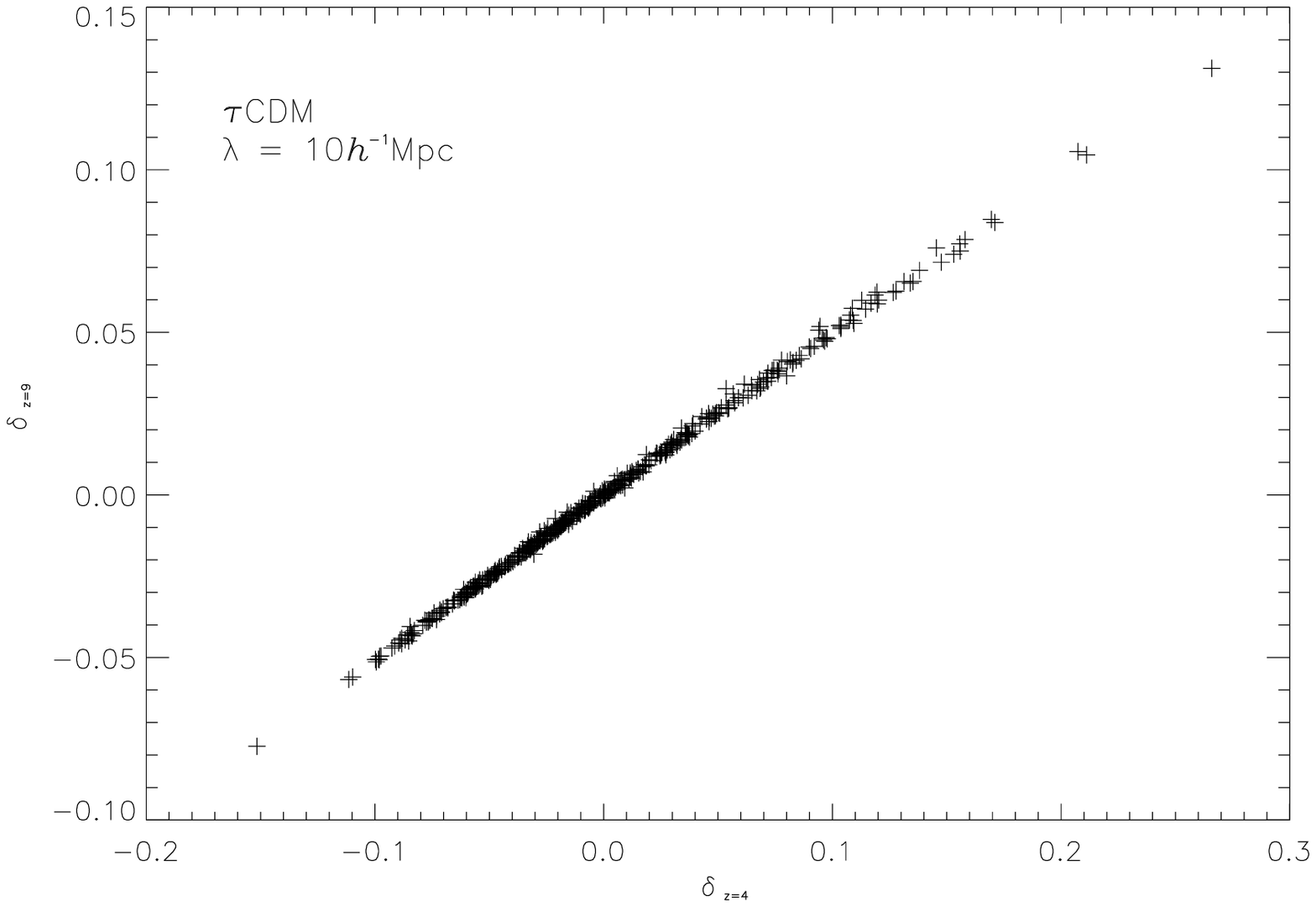}}

\caption{ Point-by-point comparison of the reconstructed $\tau$CDM density field at redshift $z=4$ and the reconstructed $\tau$CDM density field at redshift $z=9$, when the adopted characteristic smoothing lengths are of $8\lu$ and $10\lu$. \label{convscater}}
\ec
\end{figure*}

In order to test the convergence of the reconstruction method, we obtained several reconstructed density fields for diffent values of the total number of integration steps, as mentioned in \ref{recmethod}. Fig. \ref{convergence} shows a particular slice of the reconstructed density fields at different integration steps when the method is applied to the maps at $z=0$ obtained from the $\tau$CDM simulation by smoothing on characteristic lengths of $8\lu$ and $10\lu$. As it is evident from this Figure, the topology of the density fields is undistinguishable for values of $z$ greater than $4$ on smoothing scales of both $8\lu$ and $10\lu$. Only the amplitude of the density fluctuations changes, indicating that we are now in the linear regime. In Fig. \ref{convscater} this is made even clearer. Here we show a point-by-point comparison of the reconstructed $\tau$CDM density fields at redshifts $z=4$ and $z=9$, for the characteristic smoothing lengths of $8\lu$ and $10\lu$. We only show one in eight of all points, chosen randomly. It is obvious from this plot that the shape of fluctuations did not change from $z=4$ to $z=9$, for either of the smoothing lengths adopted. It is also obvious that the amplitude of fluctuations was reduced by a factor of $2$, as it is expected in the linear regime (notice that in the linear regime the growing mode of fluctuations varies as $1/(1+z)$). Hence, our time-machine shows the correct assimptotic behaviour.

\begin{figure*}
\bc
\resizebox{8cm}{!}{\includegraphics{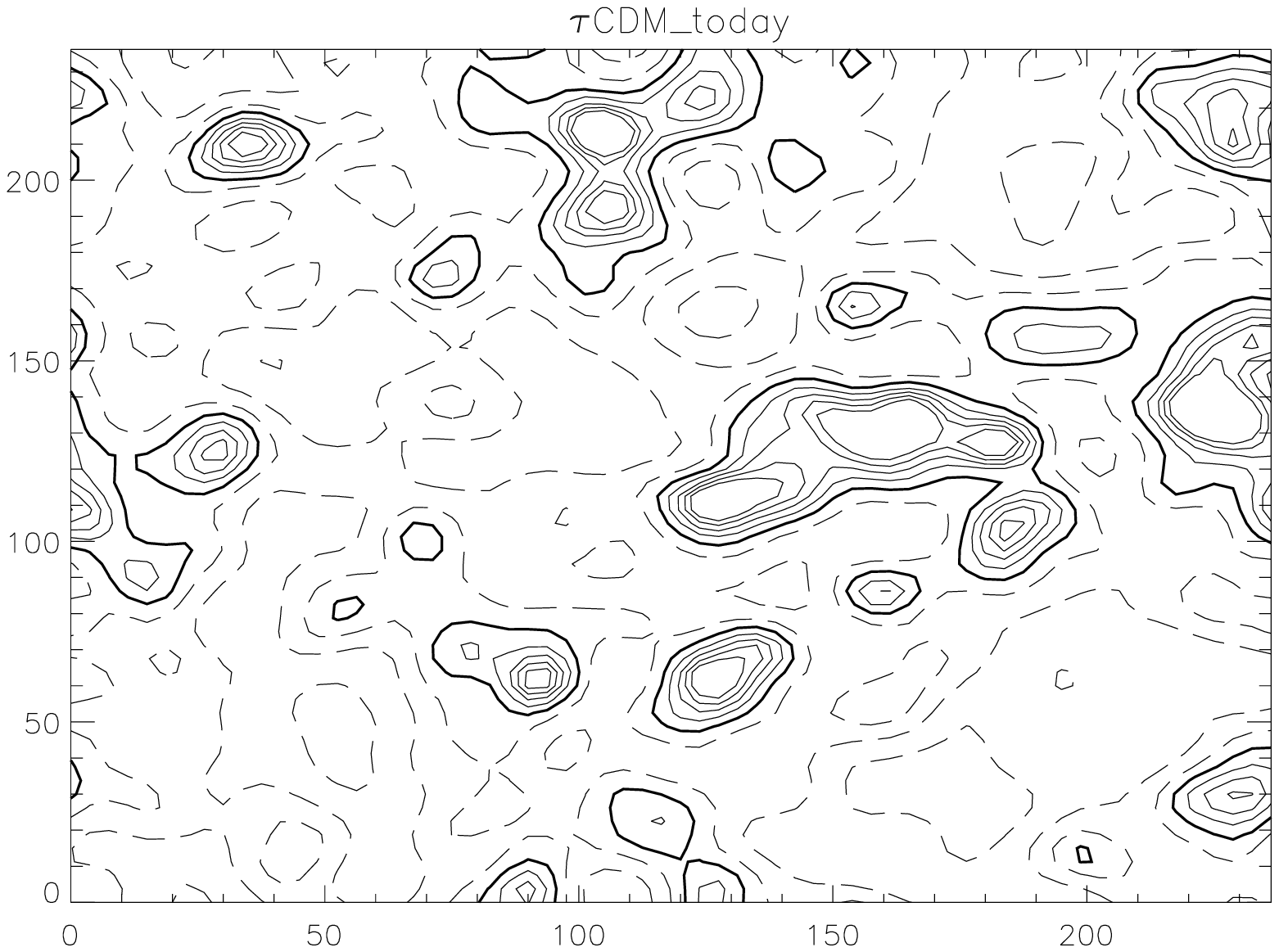}}
\resizebox{8cm}{!}{\includegraphics{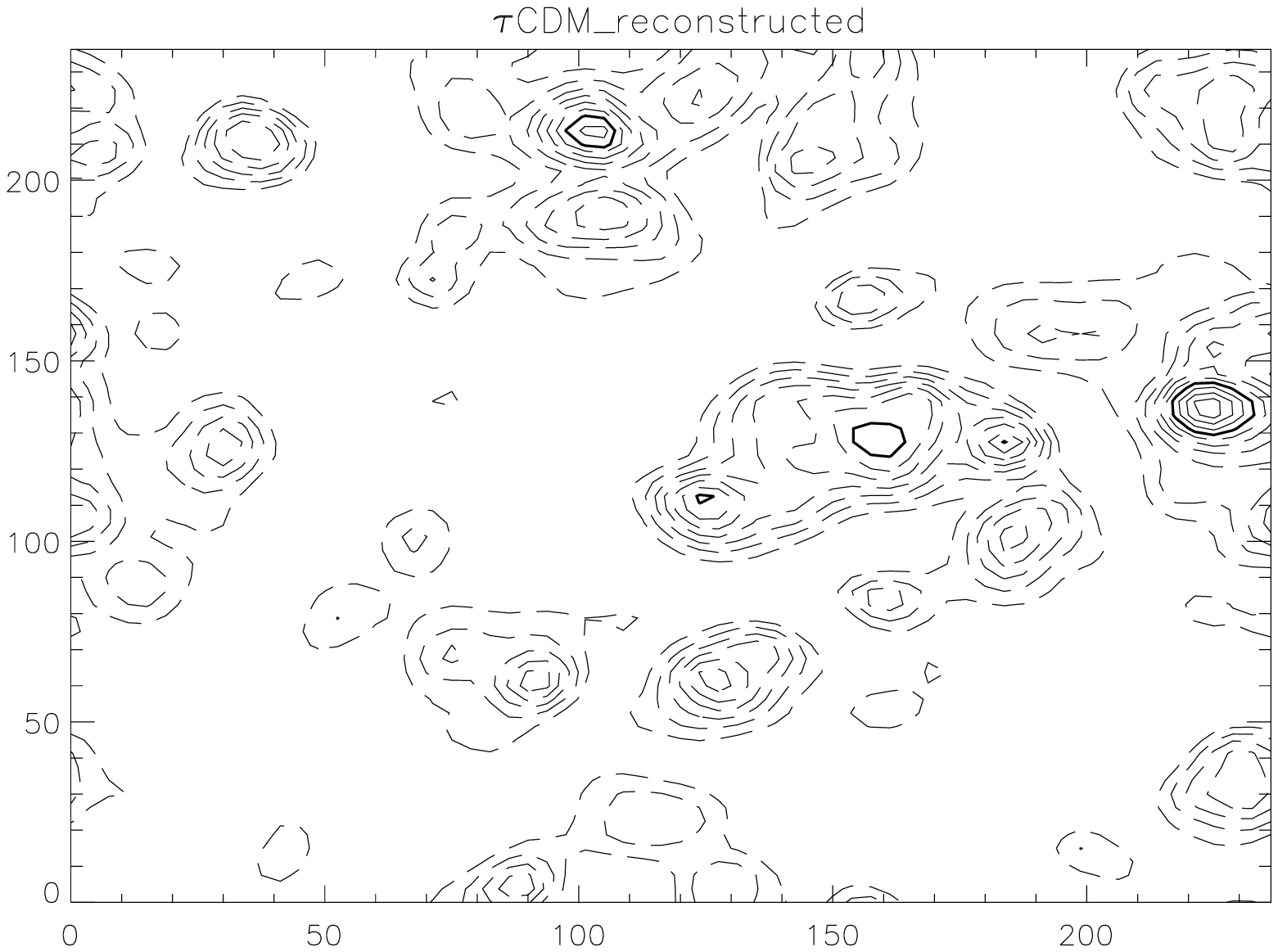}}
\resizebox{8cm}{!}{\includegraphics{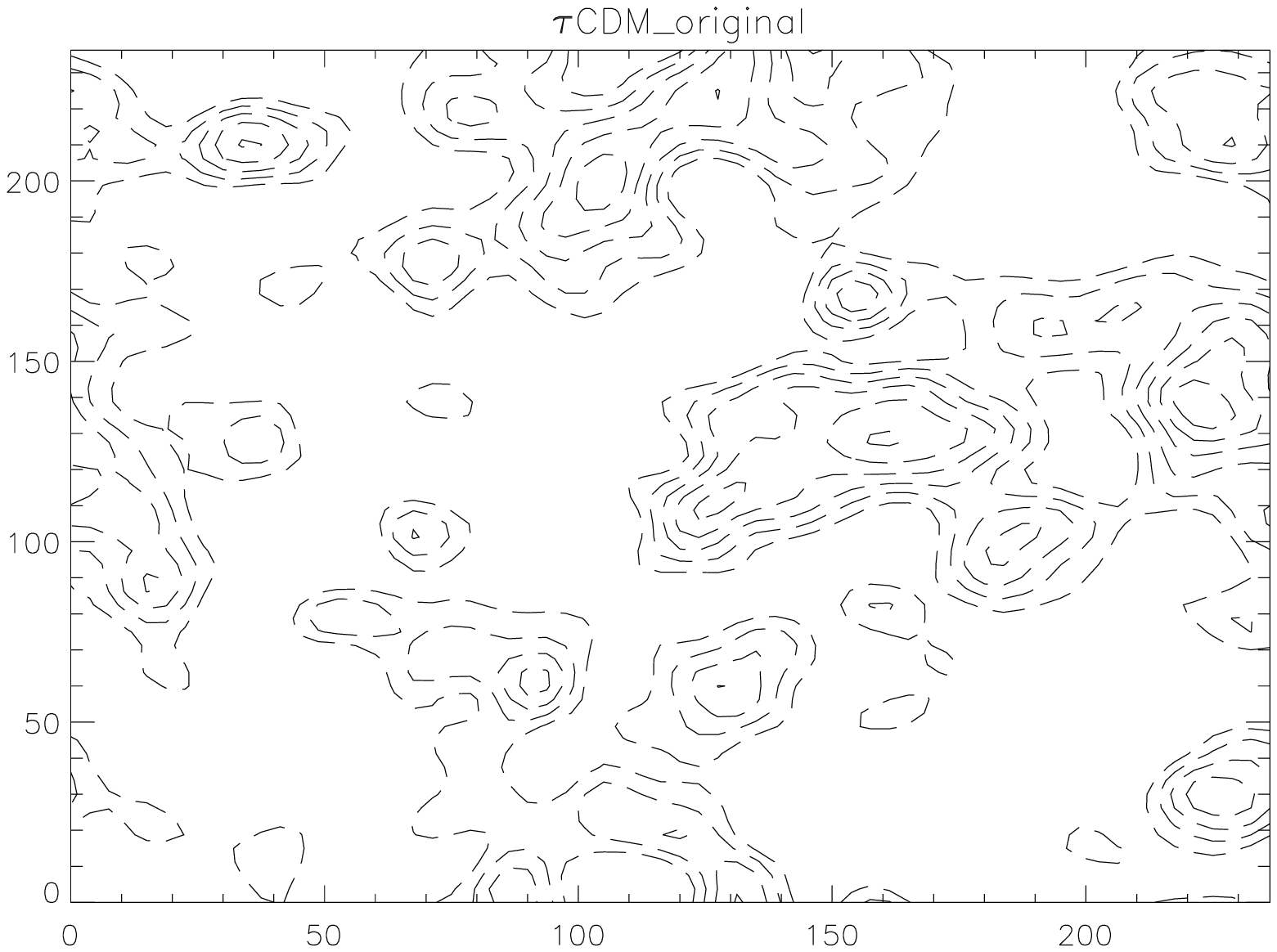}}

\caption{Isodensity contours of the $\tau$CDM simulation. The characteristic smoothing length adopted was $8\lu$. The separation between contour levels is $0.2/(1+z)$. The zero contour is heavy; positive contours are solid; negative contours are dashed.  \label{nbodycontours}}
\ec
\end{figure*}

\subsection*{Results} 

Figure \ref{nbodycontours} shows the isodensity contours obtained for the SCDM simulation and the $\tau$CDM simulation, when the fields are adaptively smoothed on a characteristic scale of $5\lu$ in a comoving box of size $240\lu$. The first row shows the fields at the present time. The second row shows the fields after applying the Zel'dovich time-machine back to a redshift of 10 and the third row shows the original fields used as input in the simulations. In each panel the thick line represents the contour where the density contrast is zero. Dashed lines represent contours of negative density contrast whereas solid lines represent positive density contrasts. The contour spacing is $0.2/(1+z)$. Because this is normalized to the linear reconstruction case where $\delta\propto 1/(1+z)$, the linear theory reconstruction density maps look exactly like the present day maps (first row), albeit their much lower fluctuation amplitude.  From the contours on both the reconstructed maps and the original maps we notice that their fluctuation amplitudes are larger than the fluctuation amplitude of a linearly reconstructed map. This result is expected because fluctuations will grow faster at the later stages of evolution, in the mildly non-linear regime. It is obvious from Fig. \ref{nbodycontours} that the Zel'dovich time-machine is able to change the rank order of isodensity contours. This means that the genus curves will also be changed. In fact we can predict, just by looking at Fig. \ref{nbodycontours} that the amplitude of the genus curves will increase after applying the Zel'dovich time-machine, as it is expected.

\begin{figure*}
\bc
\resizebox{6cm}{!}{\includegraphics{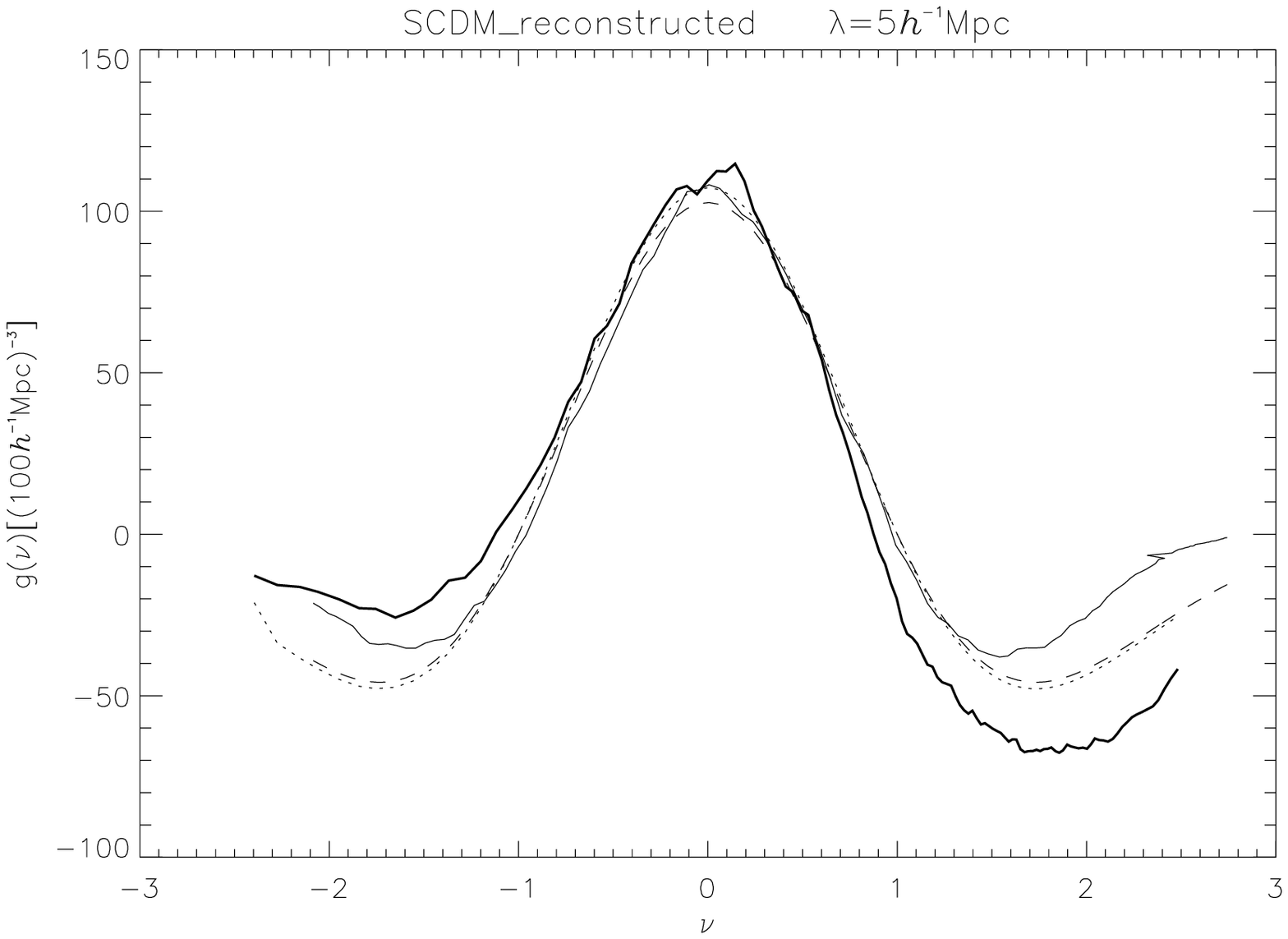}}
\resizebox{6cm}{!}{\includegraphics{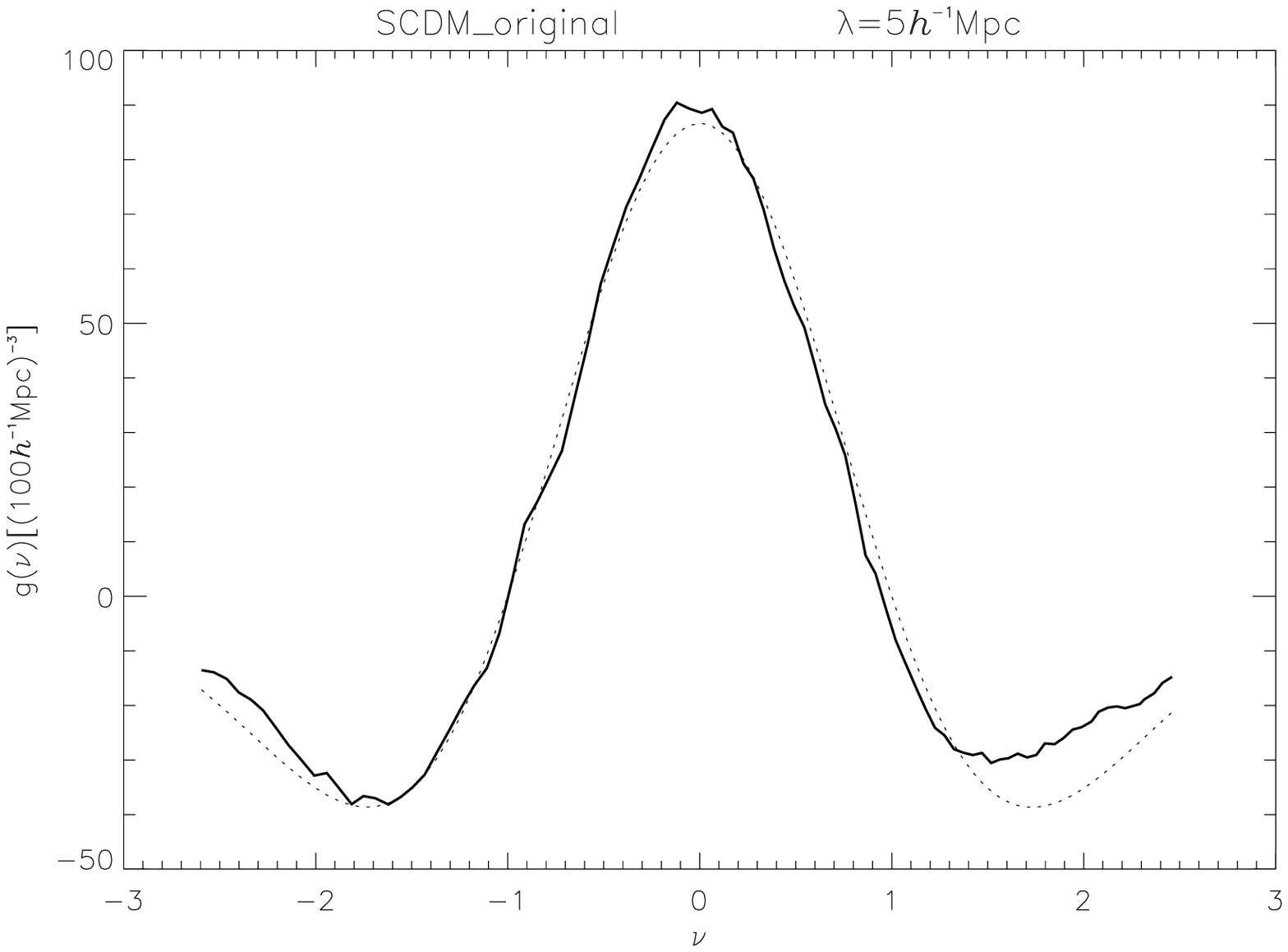}}
\resizebox{6cm}{!}{\includegraphics{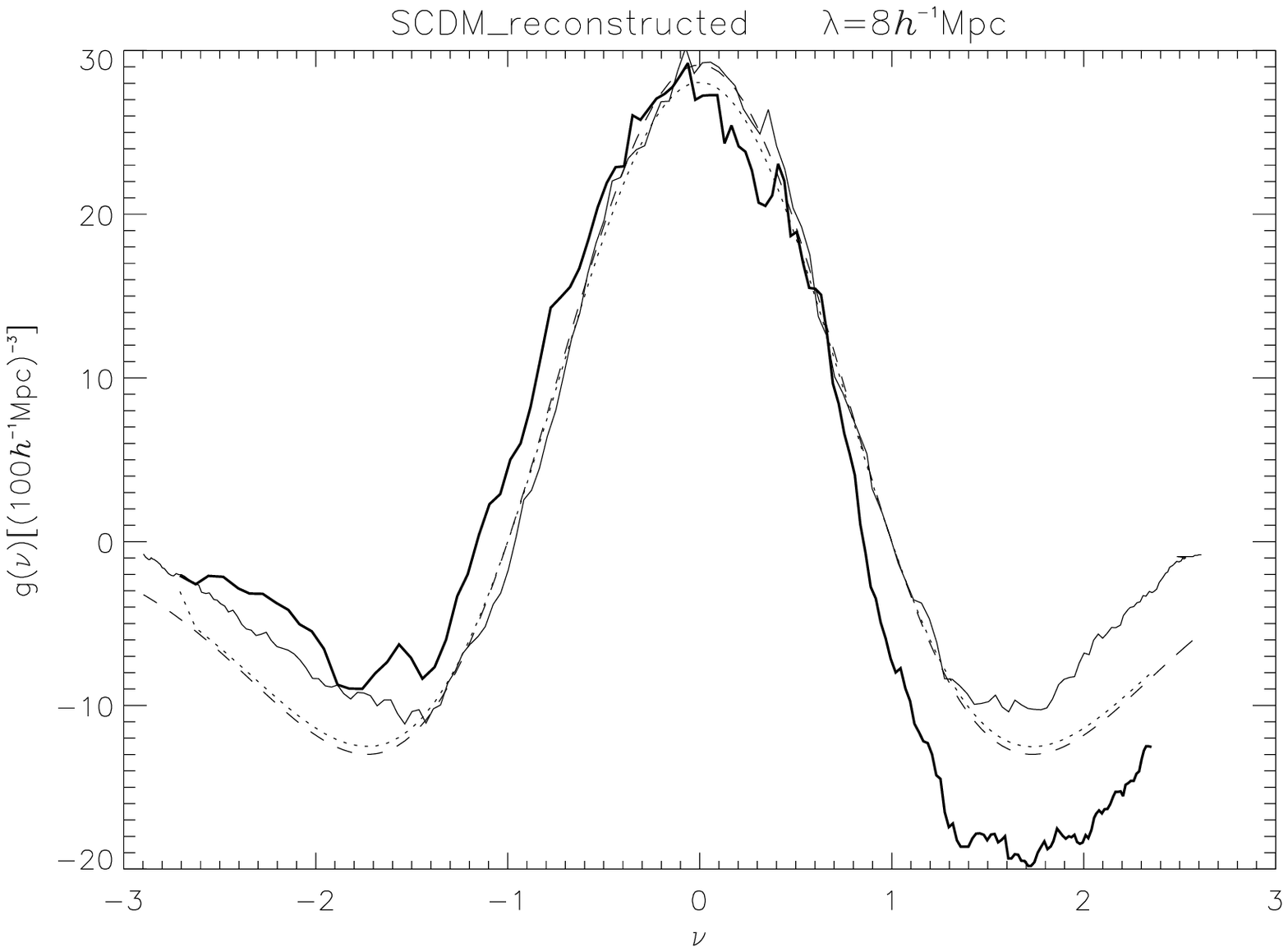}}
\resizebox{6cm}{!}{\includegraphics{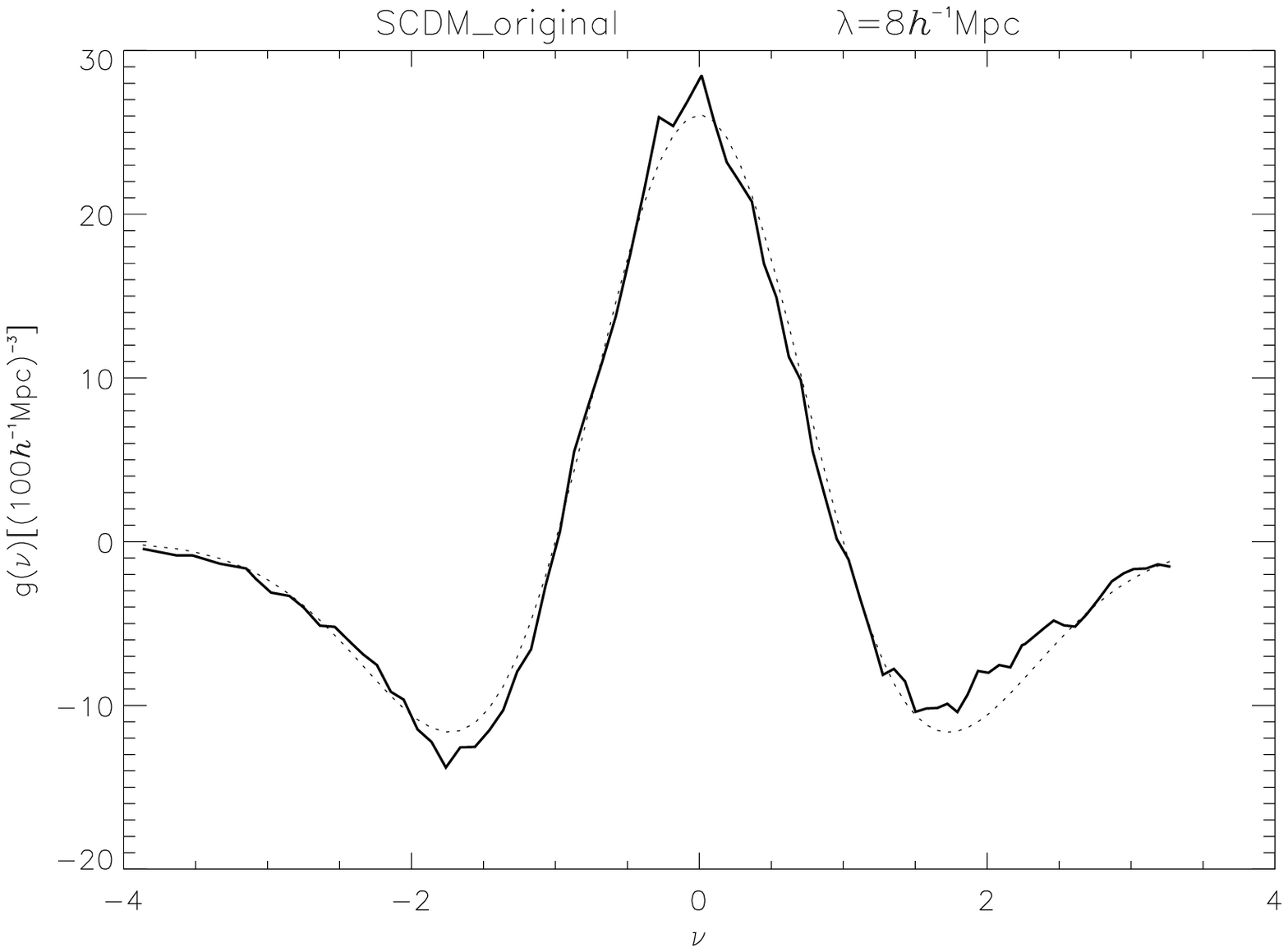}}
\resizebox{6cm}{!}{\includegraphics{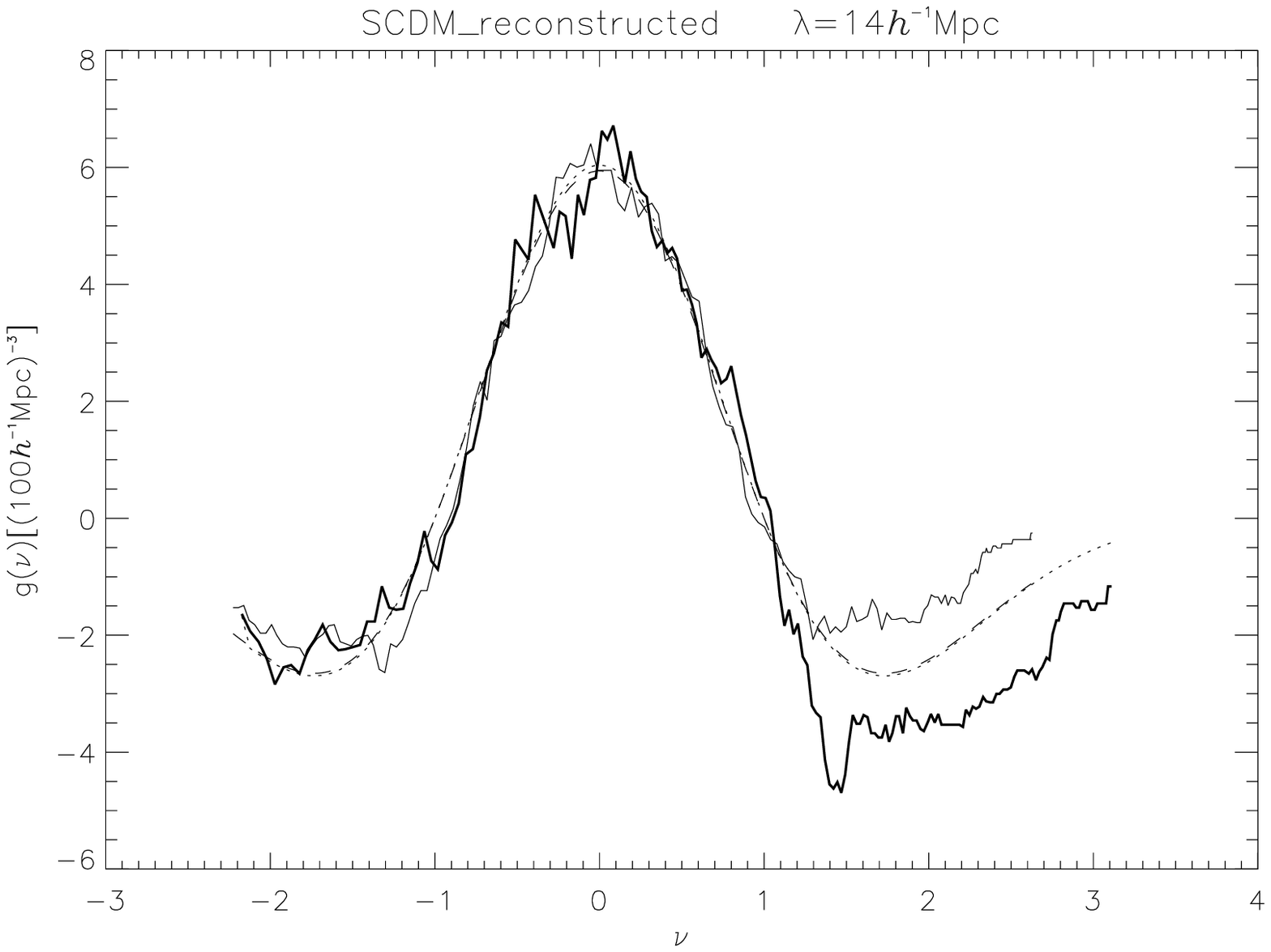}}
\resizebox{6cm}{!}{\includegraphics{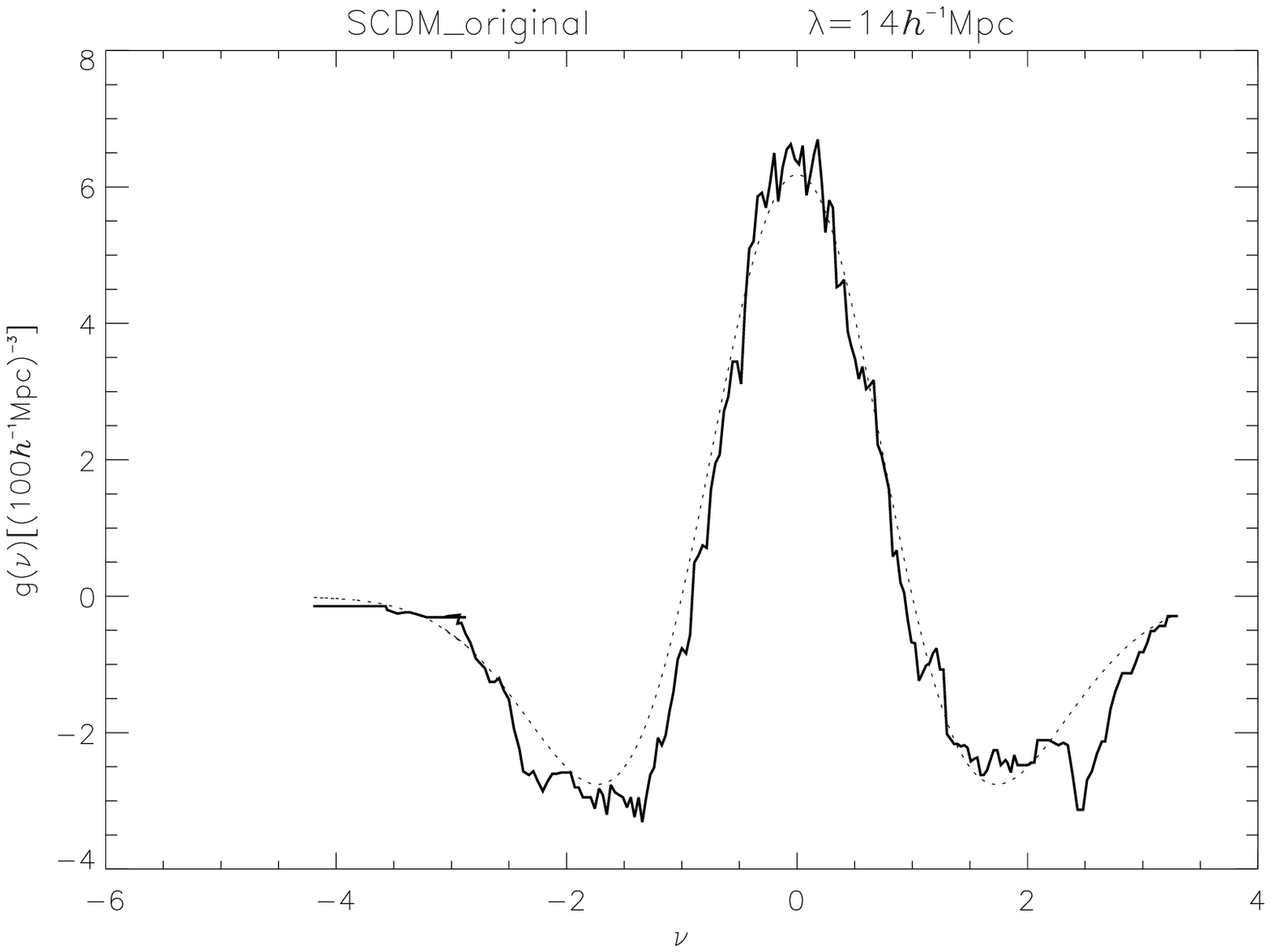}}

\caption{Genus curves for the SCDM model, at selected smoothing lengths. The left column shows the genus curves of the reconstructed field (thick solid lines) and of the randomized version of that field (thin solid lines), whereas the right column shows the genus curves of the original density field used as input for the N-body simulations. The dashed and dotted lines are the best fit random-phase curves to each of the fields. \label{SCDMgenus}}
\ec 
\end{figure*}

\begin{figure*}
\bc
\resizebox{6cm}{!}{\includegraphics{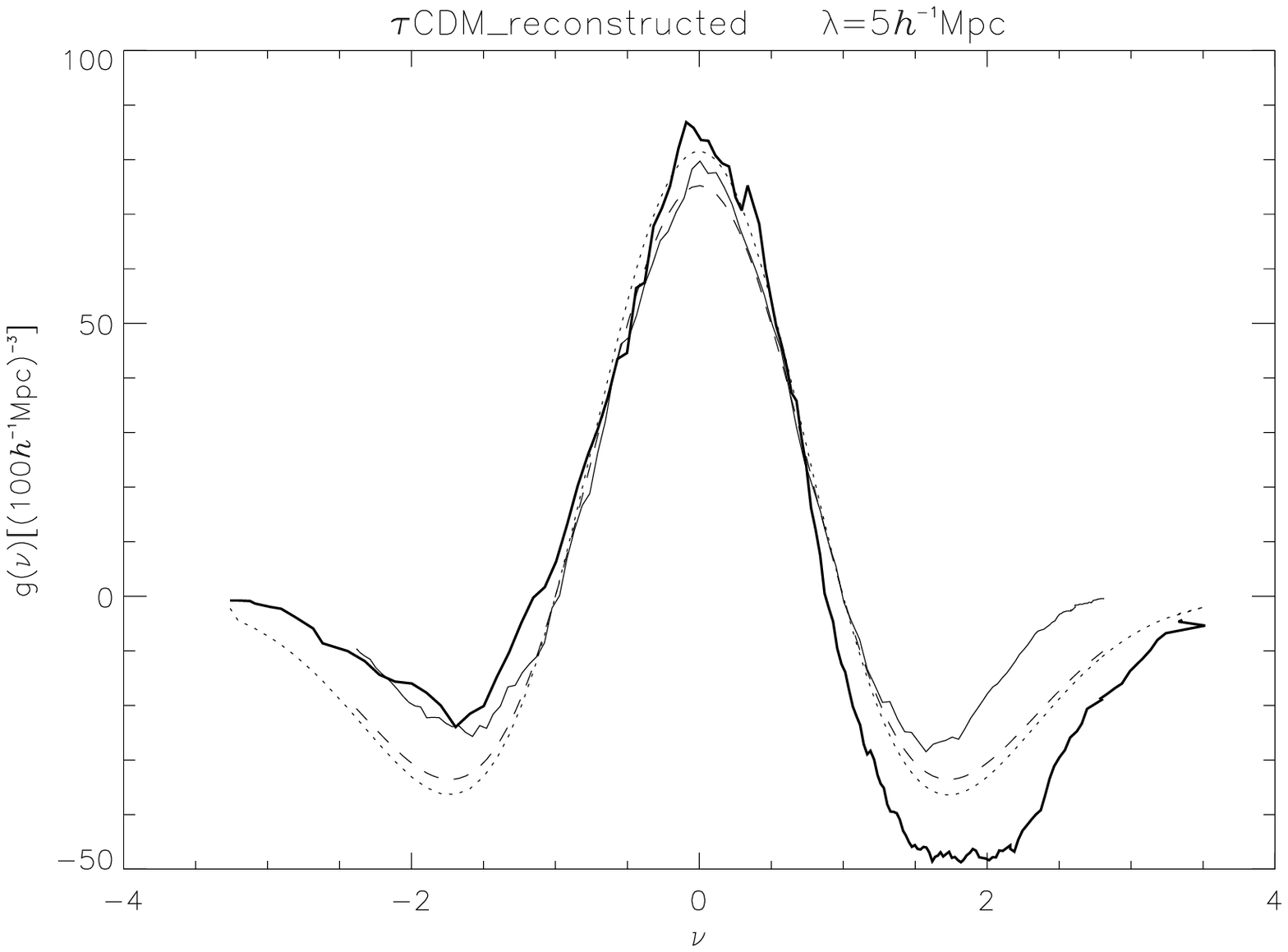}}
\resizebox{6cm}{!}{\includegraphics{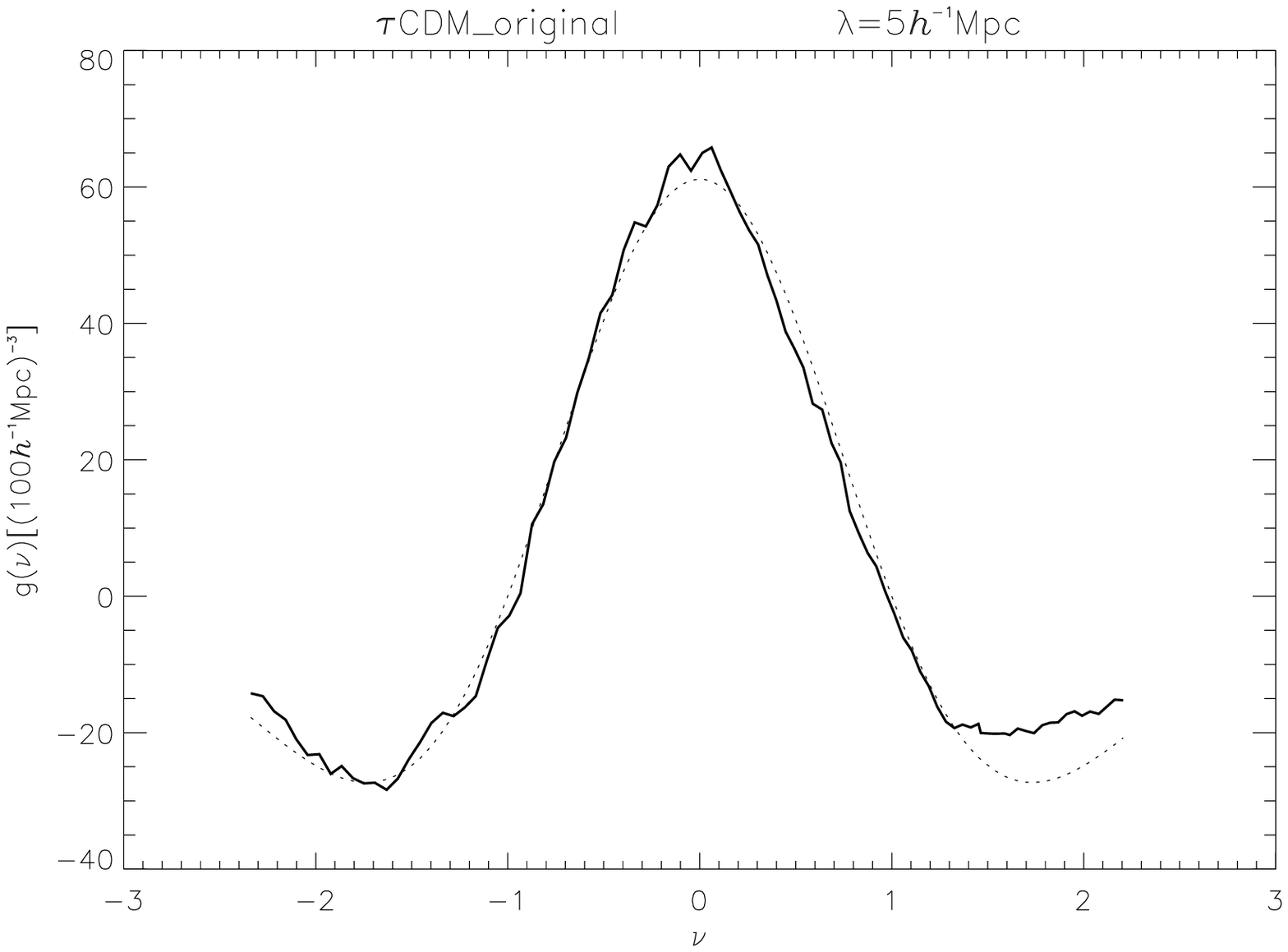}}
\resizebox{6cm}{!}{\includegraphics{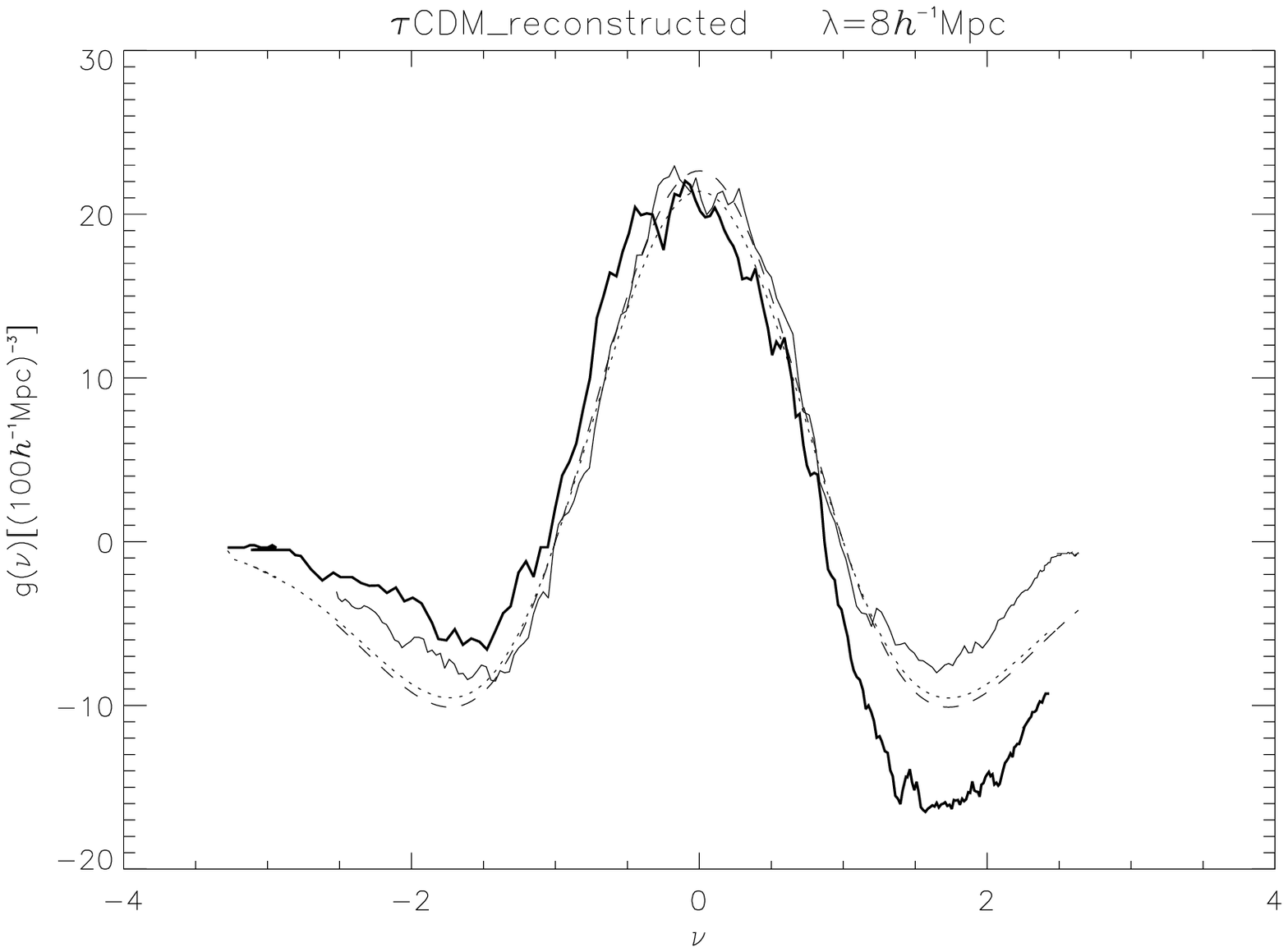}}
\resizebox{6cm}{!}{\includegraphics{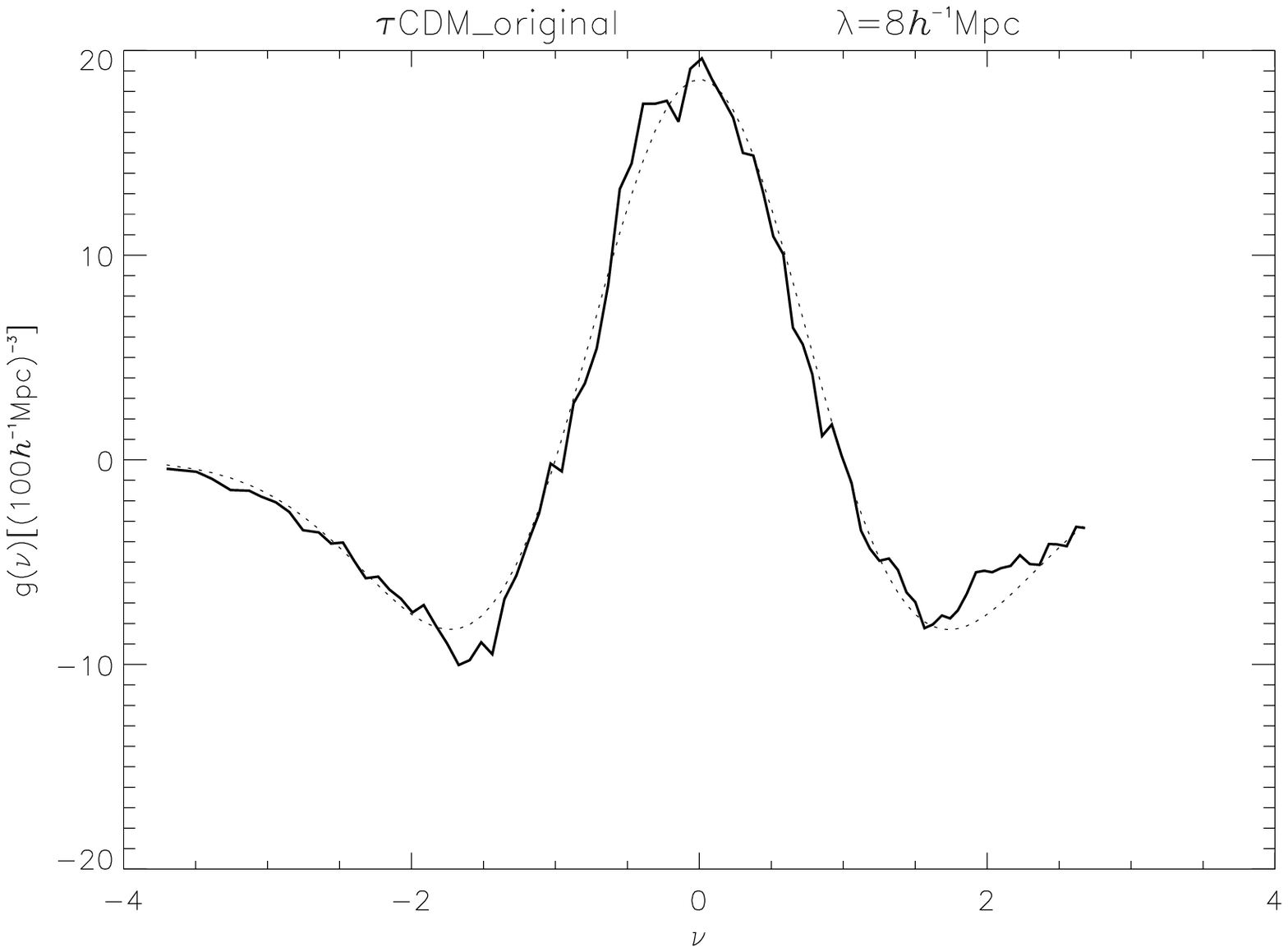}}
\resizebox{6cm}{!}{\includegraphics{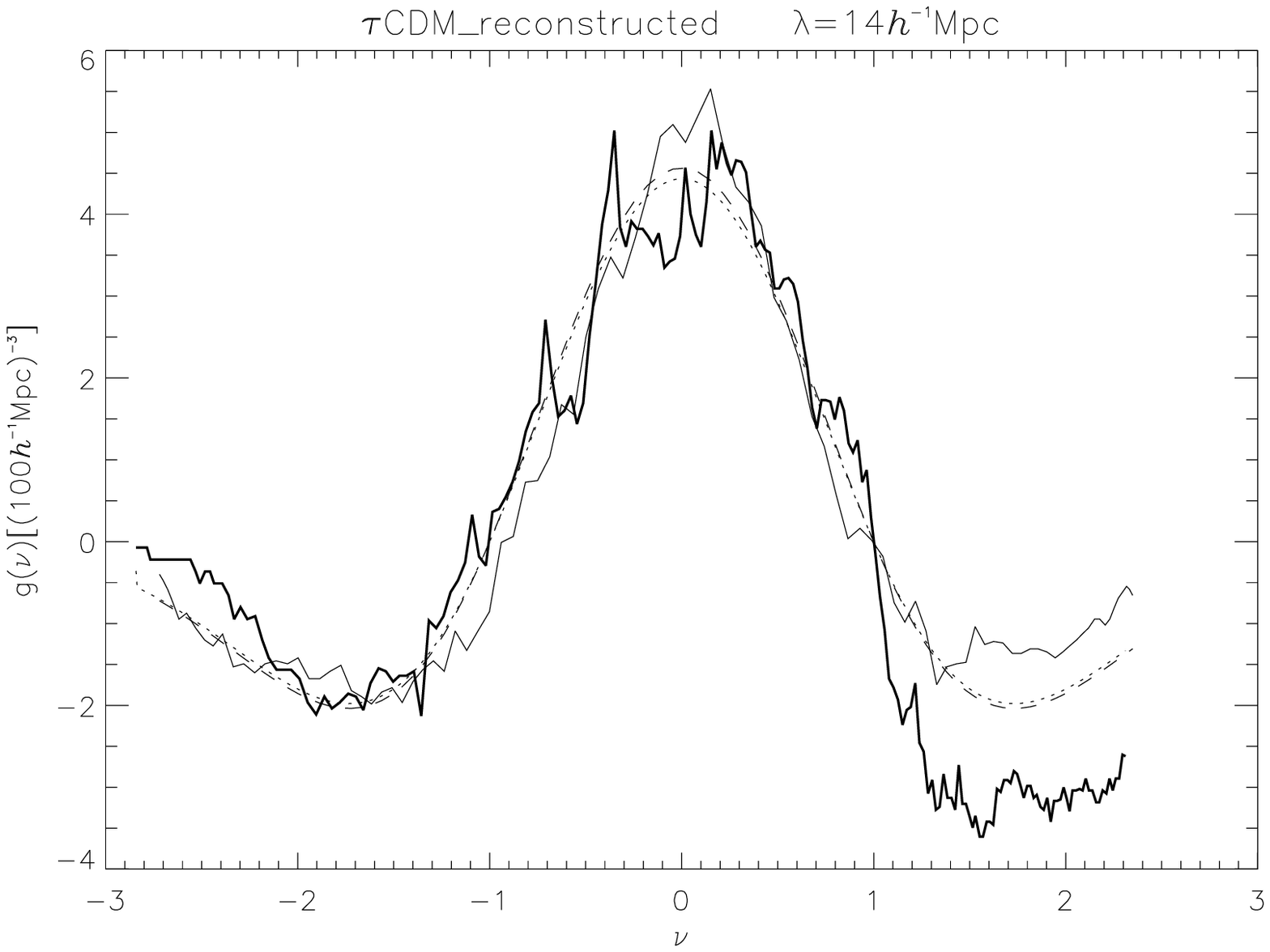}}
\resizebox{6cm}{!}{\includegraphics{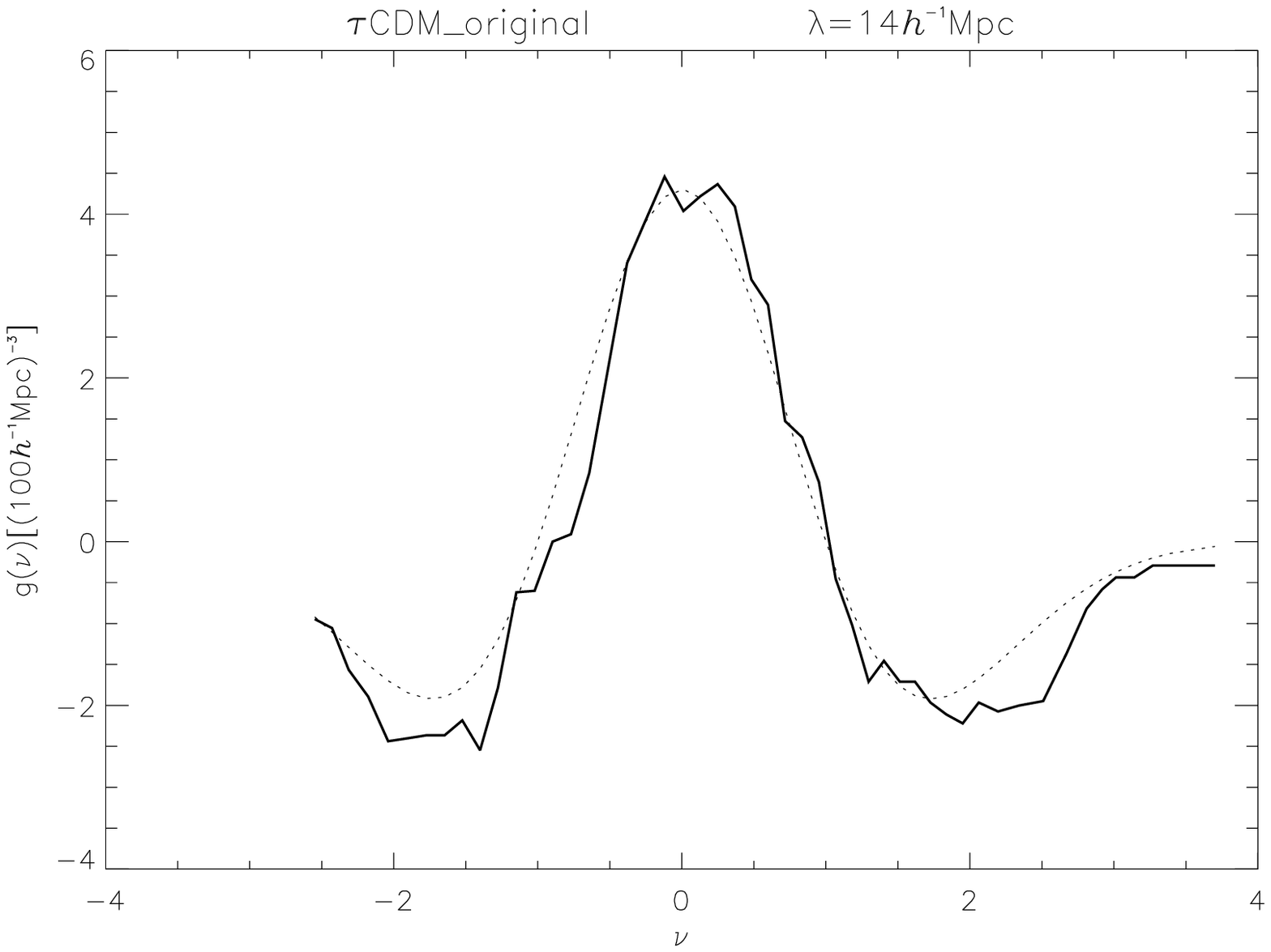}}

\caption{Genus curves for the $\tau$CDM model, at selected smoothing lengths. The left column shows the genus curves of the reconstructed field (thick solid lines) and of the randomized version of that field (thin solid lines), whereas the right column shows the genus curves of the original density field used as input for the N-body simulations. The dashed and dotted lines are the best fit random-phase curves to each of the fields.\label{TCDMgenus}}
\ec
\end{figure*}
\begin{figure}
\bc
\resizebox{8cm}{!}{\includegraphics{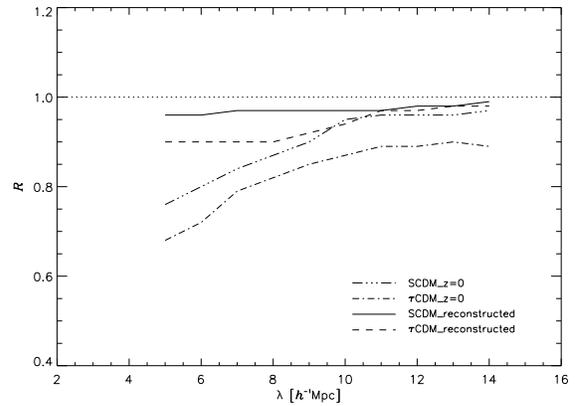}}
\caption{Amplitude drops measured for the reconstructed SCDM simulation and $\tau$CDM simulation. Also show for comparison are the amplitude drops measured for the simulations at $z=0$\label{recdropbody}.}
\ec
\end{figure}
\begin{figure}
\bc
\resizebox{8cm}{!}{\includegraphics{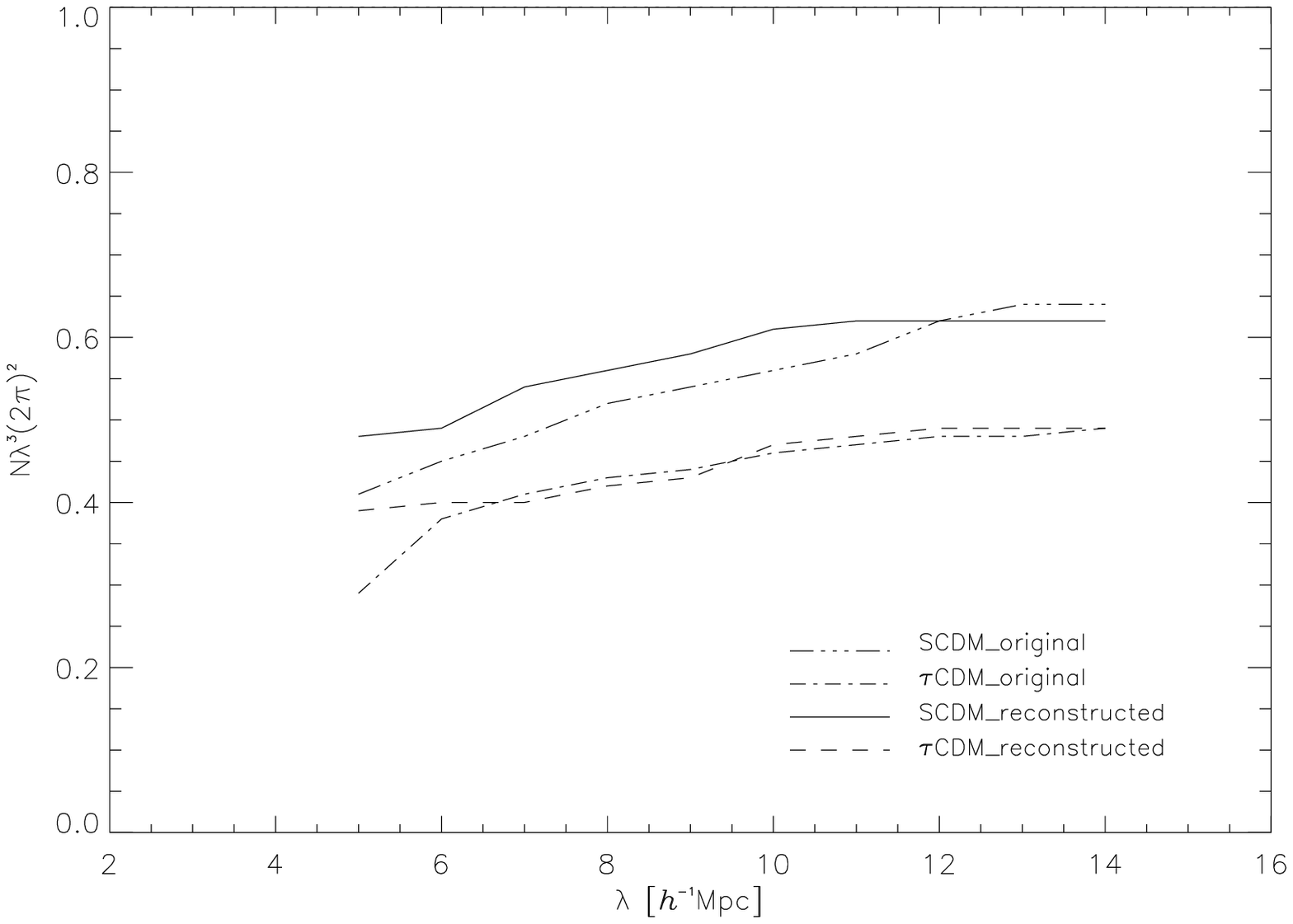}}
\caption{Genus amplitudes measured for the reconstructed SCDM simulation and $\tau$CDM simulation. Also shown are the genus amplitudes of the original fields\label{recampbody}.}
\ec
\end{figure}

Figures \ref{SCDMgenus} and \ref{TCDMgenus} show the genus curves obtained for the SCDM model and the $\tau$CDM model respectively, at selected characteristic smoothing lengths, namely at $5\lu$, $8\lu$ and $14\lu$. We restrict our analysis to this range of smoothing lengths, as we know {\em a priori} that the Zel'dovich time-machine is only relevant in the mildly non-linear regime. In both figures, the first column shows the genus curves of the reconstructed field and its randomized counterpart, whereas the second column shows the genus curves of the original fields, used as input for the simulations. We deliberately do not show the genus curves of the present day {\em adaptively smoothed} density fields, since, as we argued in the previous chapter, the isodensity contours are, at this stage, different from the isodensity contours of the fields smoothed using a constant kernel.

It is striking in Fig. \ref{SCDMgenus} and Fig. \ref{TCDMgenus} the consistency between the genus curves of the reconstructed field and of its randomized counterpart. This is particularly true for the genus amplitudes, which indicates that there is very little phase correlation in the reconstructed fields on all scales, as it is expected from Gaussian initial conditions. With regard to the genus amplitudes {\em per se}, we find consistency with the genus amplitudes of the original density fields, although we notice a tendency for these to be slightly higher in the reconstructed fields. These means that the second moment of the power spectrum is recovered slightly {\em in excess}. This becomes more important as we approach the smaller scales, which is to say as the system becomes more non-linear.

It is also interesting to notice a depression at high values of $\nu$ on the genus curves of the reconstructed fields. We find this depression to be present in all cases, both for the Virgo simulations and the PSCz data. We expect this to be a particular feature of the reconstruction technique, i.e., ultimately dependent on the Zel'dovich approximation itself. At high values of $\nu$ the regime is considerably non-linear and so we expect the Zel'dovich approximation to perform rather poorly. This problem is enhanced by the fact that the density fields have been smoothed adaptively. Nevertheless, we do not expect this feature to affect our results considerably as the calculation of the amplitude of the genus curves is restricted to the range $-1<\nu<1$.

Note that there is no sampling noise in these density fields. The tremble in the genus curves is ultimately due to cosmic variance.

In Fig. \ref{recdropbody} we plot the amplitude drops obtained for the reconstructed SCDM model and for the reconstructed $\tau$CDM model as a function of smoothing length. Also shown are the amplitude drops obtained for the models at $z=0$ \cite{Ca98}. In all scales, the values obtained for the reconstructed fields are closer to unity than those of the present-day fields. This indicates that the Zel'dovich time-machine is a good tool in recovering Gaussian initial density fields. However, at small scales, we detect a slight departure from the expected value of unity, even for the reconstructed fields. This is not surprising since at this scales non-linearities become strong. The degree of strength of the non-linearities depends on the particular model. As we see from Fig. \ref{recdropbody} they are more important in the $\tau$CDM model than in the SCDM model, in agreement to what was found by Canavezes et al. \shortcite{Ca98}.

Fig. \ref{recampbody} shows the genus amplitudes of the reconstructed  fields for both the SCDM model and the $\tau$CDM model, together with the genus amplitudes of the original fields. In the case of the $\tau$CDM model the agreement between the original genus amplitude and the genus amplitude of the reconstructed field, is striking, in particular when we consider scales above $\sim 7\lu$. For the SCDM model the agreement is not as good. The reconstructed genus amplitudes appear to be slightly higher than the true original ones.

These results indicate that the Zel'dovich time-machine is effective in recovering the right genus amplitude {\em drops} on scales larger than $\sim 8\lu$, although recovered amplitudes cannot be considered reliable in the strict sense of the word.

\section{RECOVERING THE INITIAL DENSITY FIELD FROM PSCZ}
\label{recpscz}

\begin{figure*}
\bc
\resizebox{8cm}{!}{\includegraphics{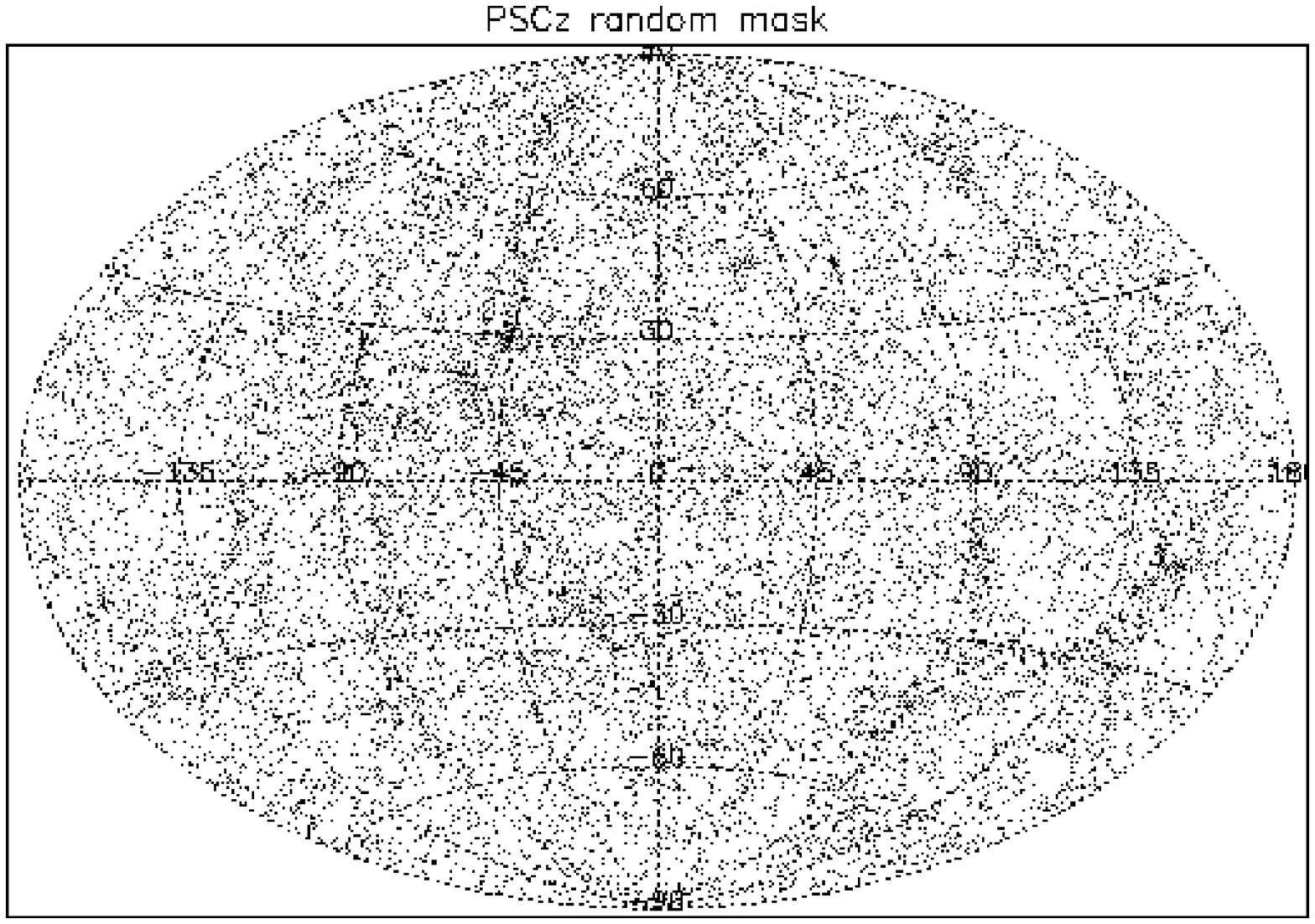}}
\resizebox{8cm}{!}{\includegraphics{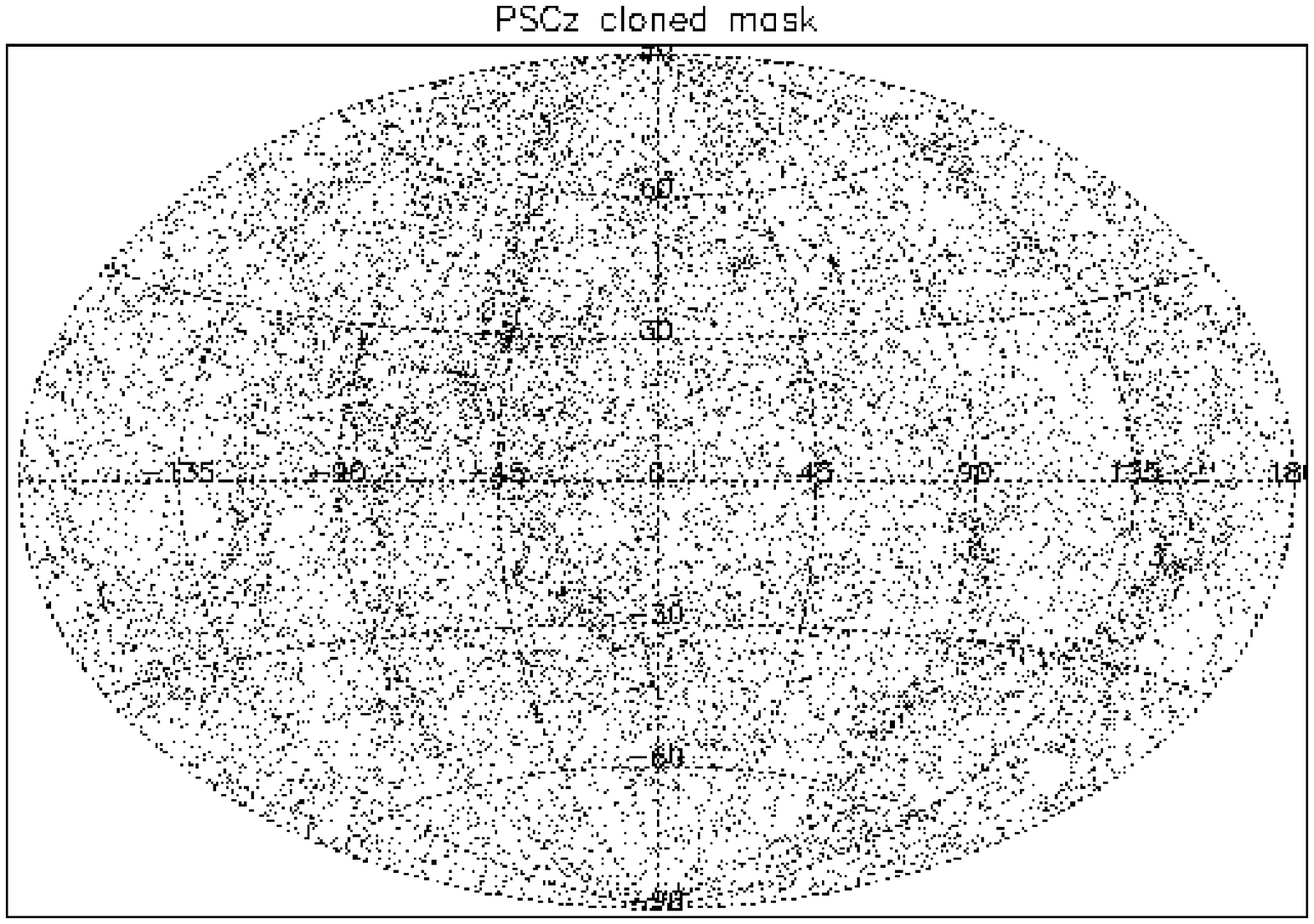}}
\caption{Aitoff projections of PSCz galaxies and filled masks.In the upper panel the mask is filled randomly whereas in the lower panel the mask is filled using a cloning technique.\label{randcloned}}
\ec
\end{figure*}

As mentioned previously, one of the difficulties in using the PSCz in an attempt to recover the initial density fluctuations in the Universe, is the fact that the masked region can make up to $\sim 20\%$ of the whole observable area. Since this region will have a definite gravitational influence on the observable area when we attempt to reconstruct to original density field, some sort of filling is essential. As mentioned in \ref{recdenmaps} we employ two different techniques: The random filling and the cloning. Fig. \ref{randcloned} shows how the two different fillings appear to the eye. On the upper panel we show the map where the mask has been filled randomly and on the lower panel we show the map where the mask has been cloned. Although it is impossible to the eye to recognize any significant difference, we will be able to detect some noticeable differences on the topologies of the reconstructed density fields.

\begin{figure*}
\bc
\resizebox{6cm}{!}{\includegraphics{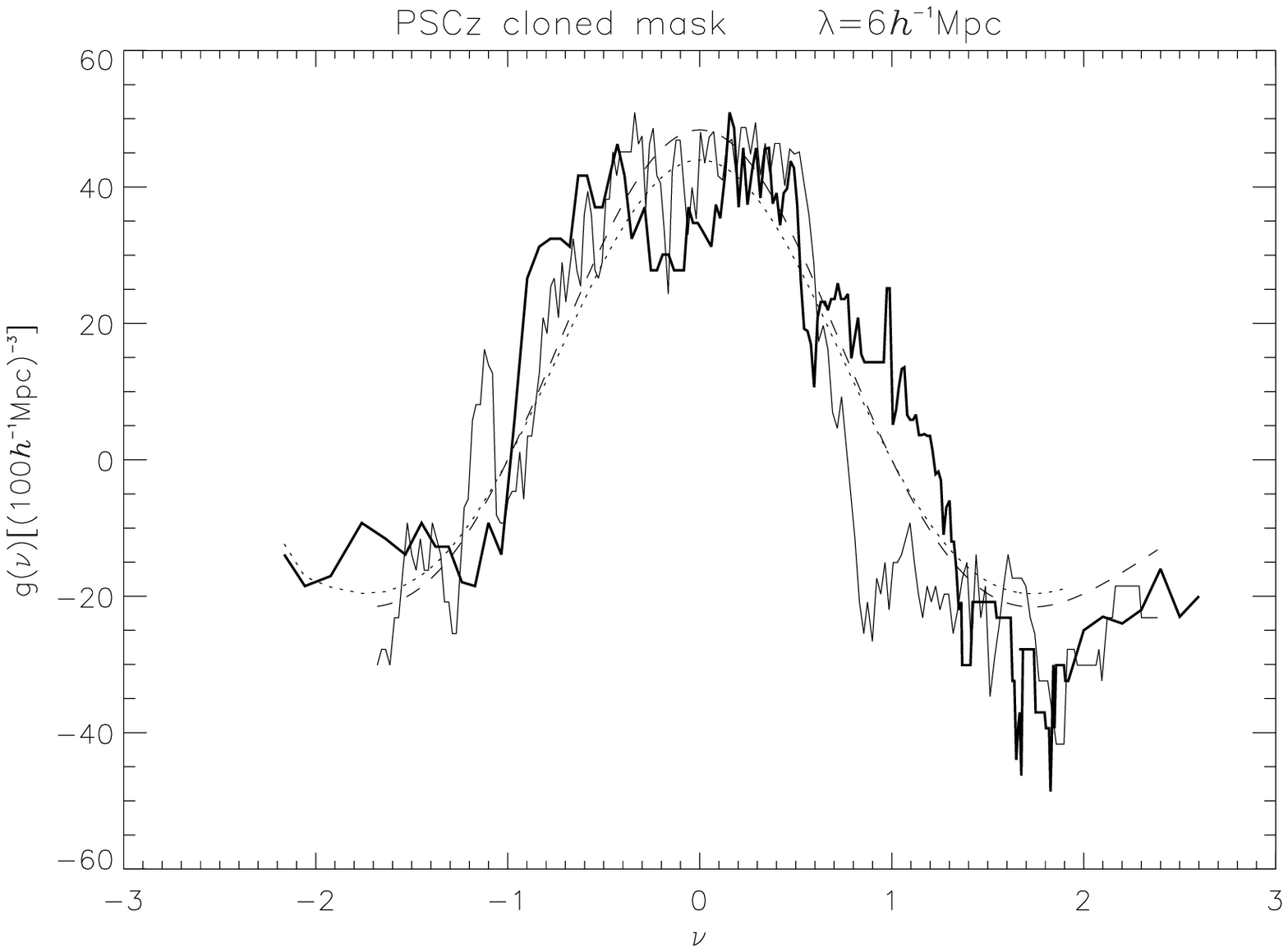}}
\resizebox{6cm}{!}{\includegraphics{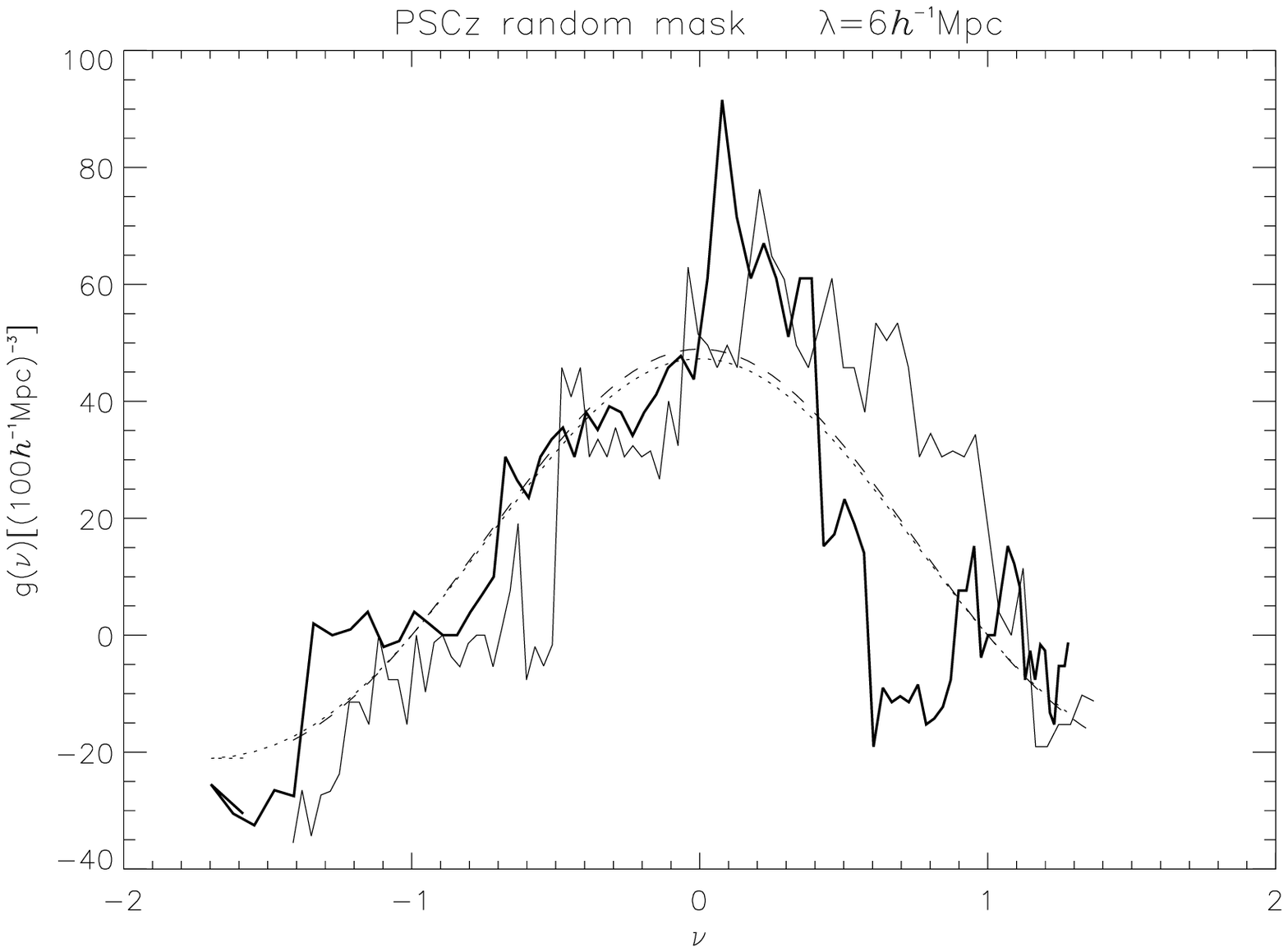}}
\resizebox{6cm}{!}{\includegraphics{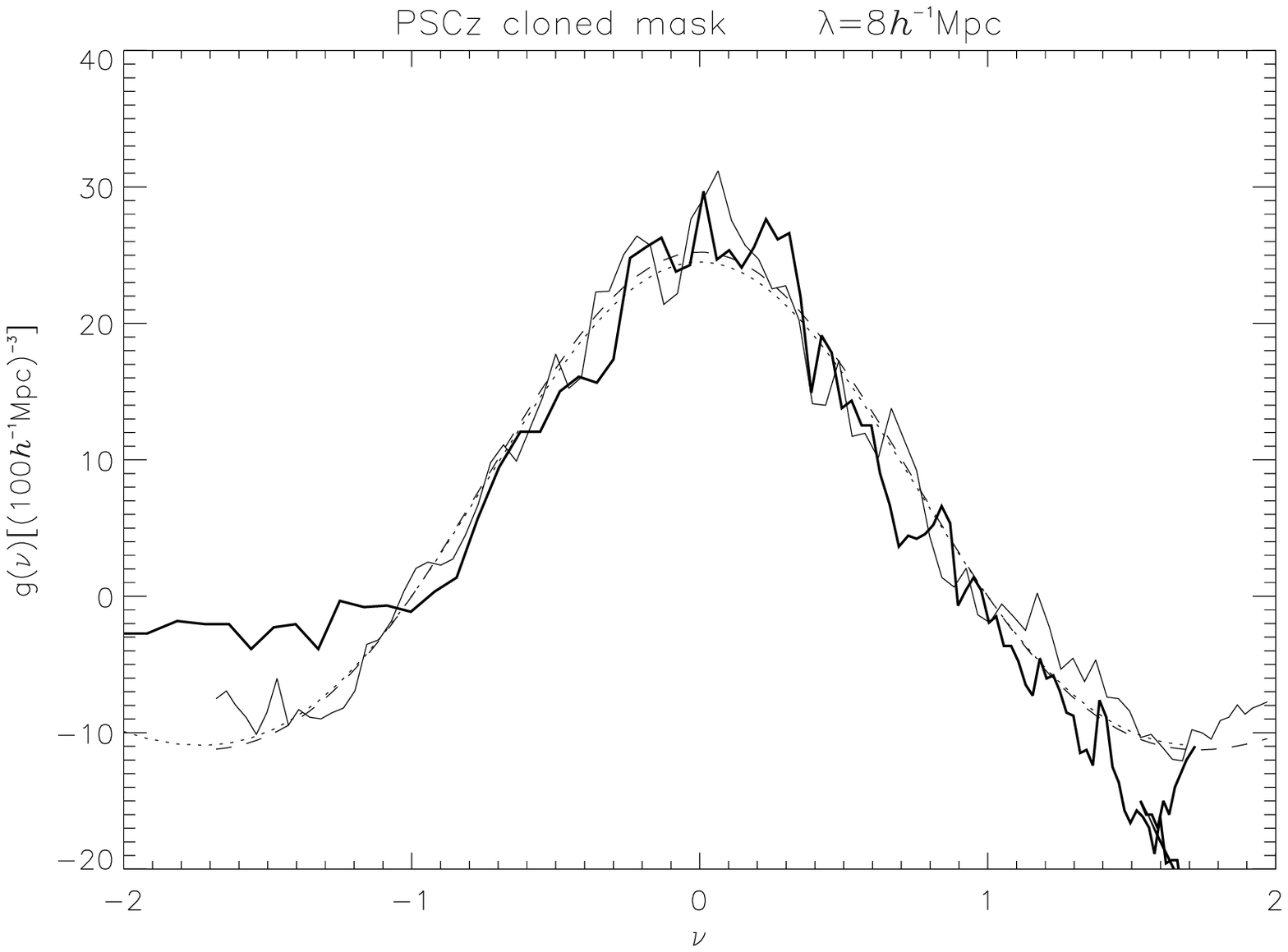}}
\resizebox{6cm}{!}{\includegraphics{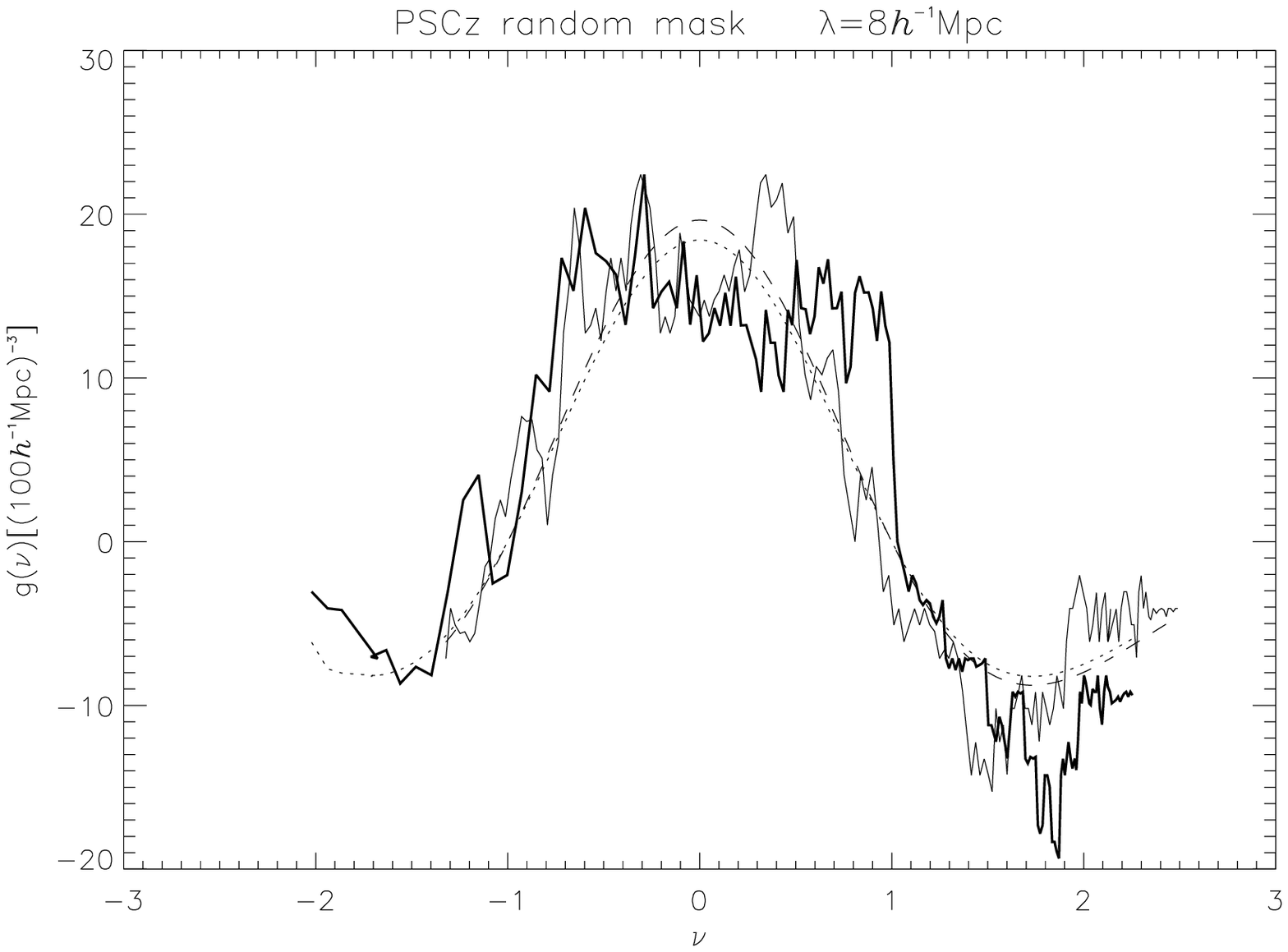}}
\resizebox{6cm}{!}{\includegraphics{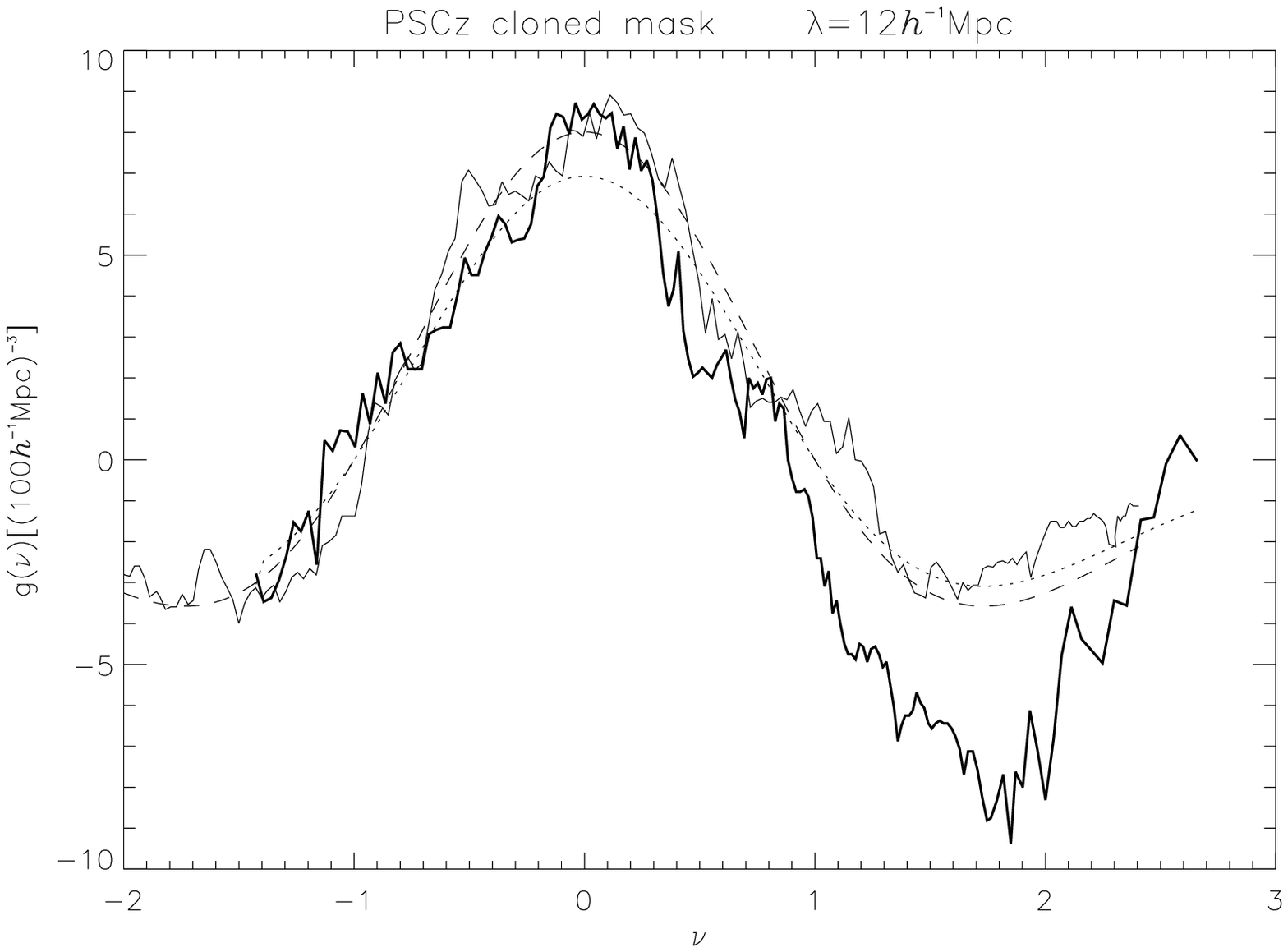}}
\resizebox{6cm}{!}{\includegraphics{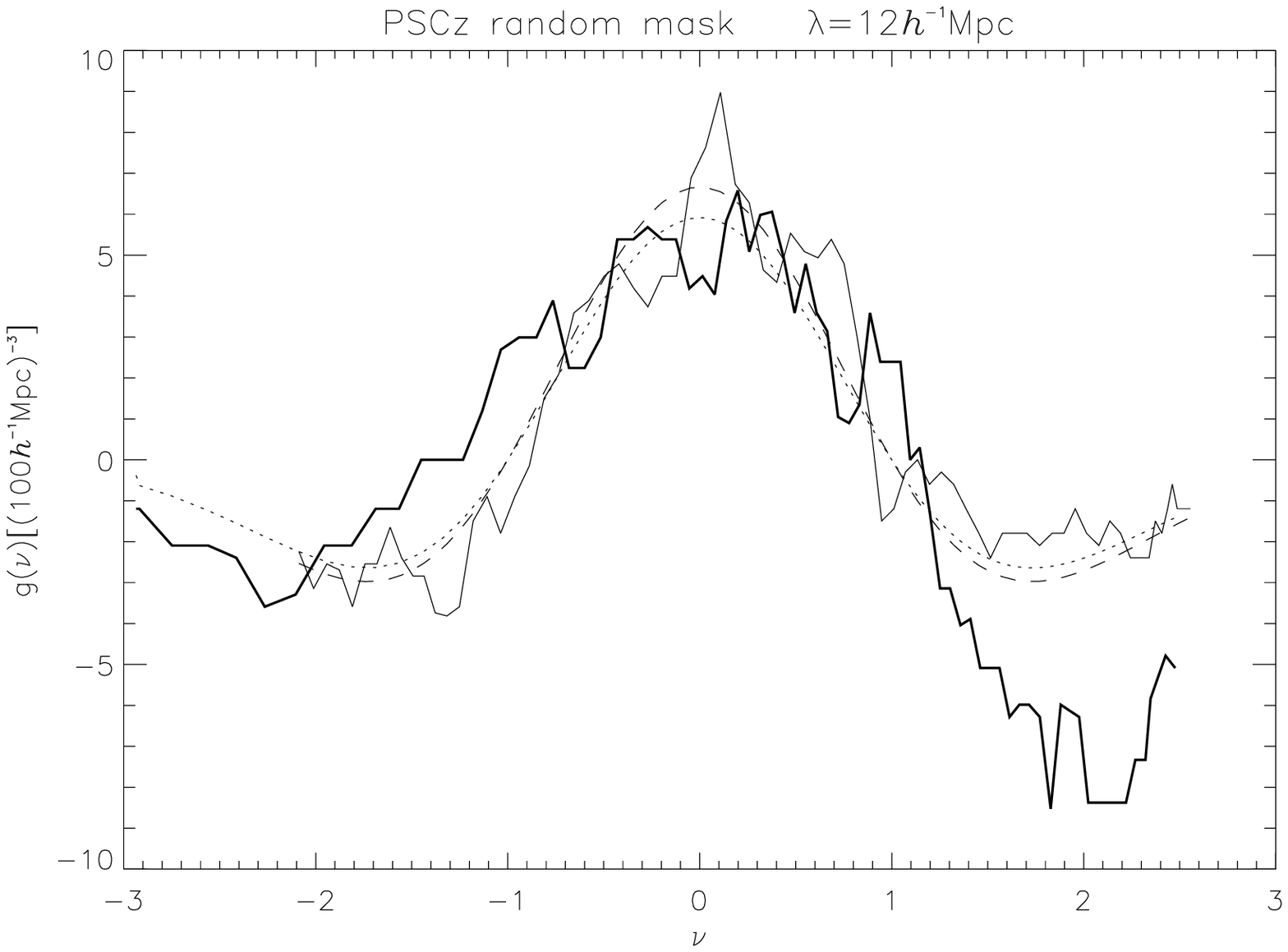}}
\resizebox{6cm}{!}{\includegraphics{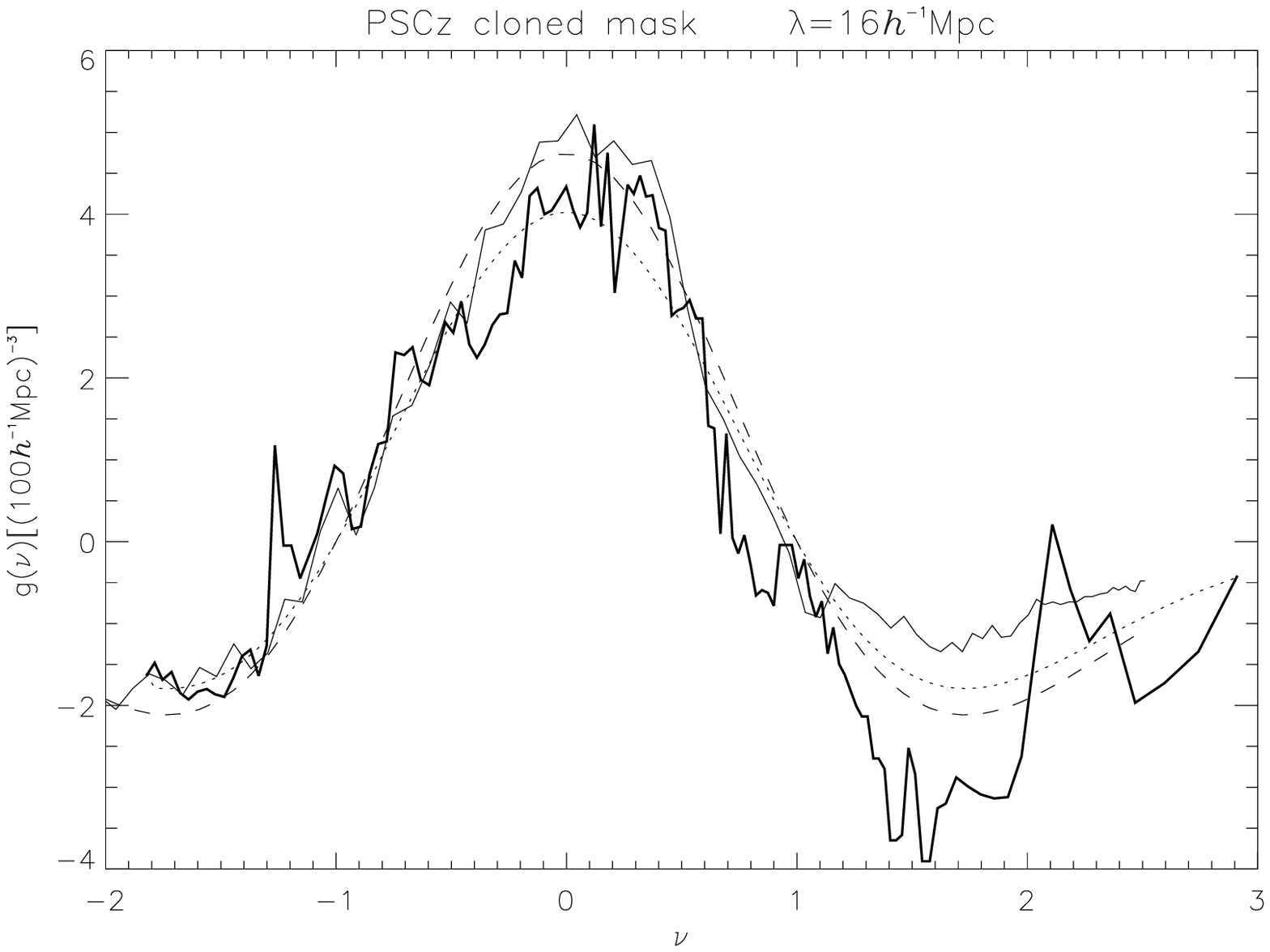}}
\resizebox{6cm}{!}{\includegraphics{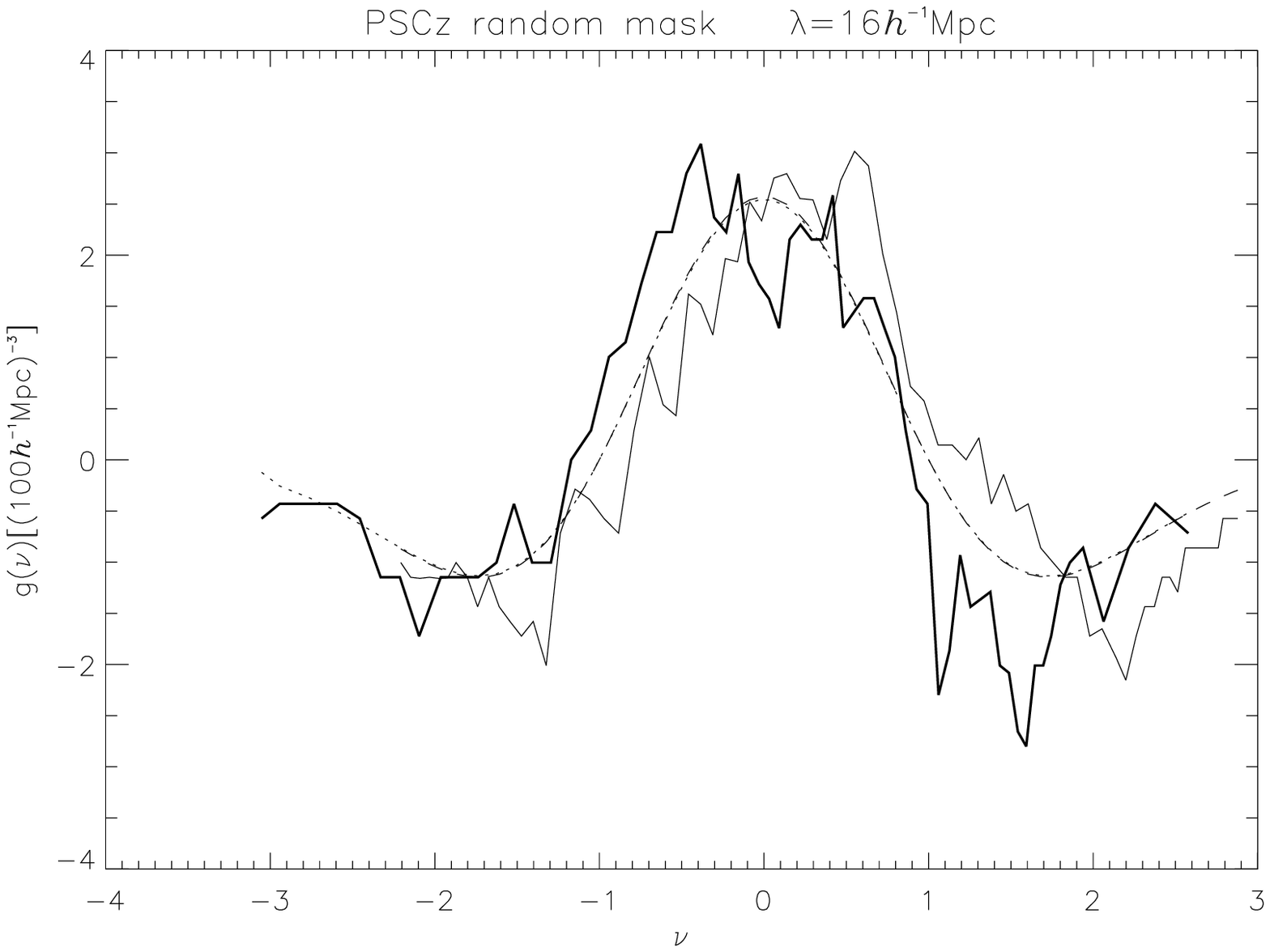}}
\caption{Genus curves of the reconstructed PSCz fields. \label{PSCzgenus}}
\ec
\end{figure*}
\begin{figure*}
\bc
\resizebox{10cm}{!}{\includegraphics{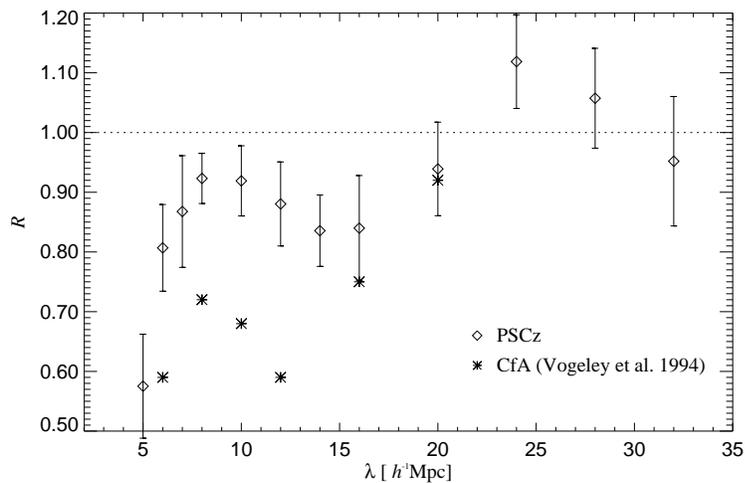}}
\caption{The amplitude drop $R$ measured for PSCz. Also shown are the results
obtained by Vogeley et al.$1994$ for the CfA survey. Here the 
errors are taken to be the uncertainty Vogeley et al.$1994$ report 
for mock catalogues extracted from a LCDM model\label{drpscznow}}
\ec
\end{figure*}
\begin{figure*}
\bc
\resizebox{8cm}{!}{\includegraphics{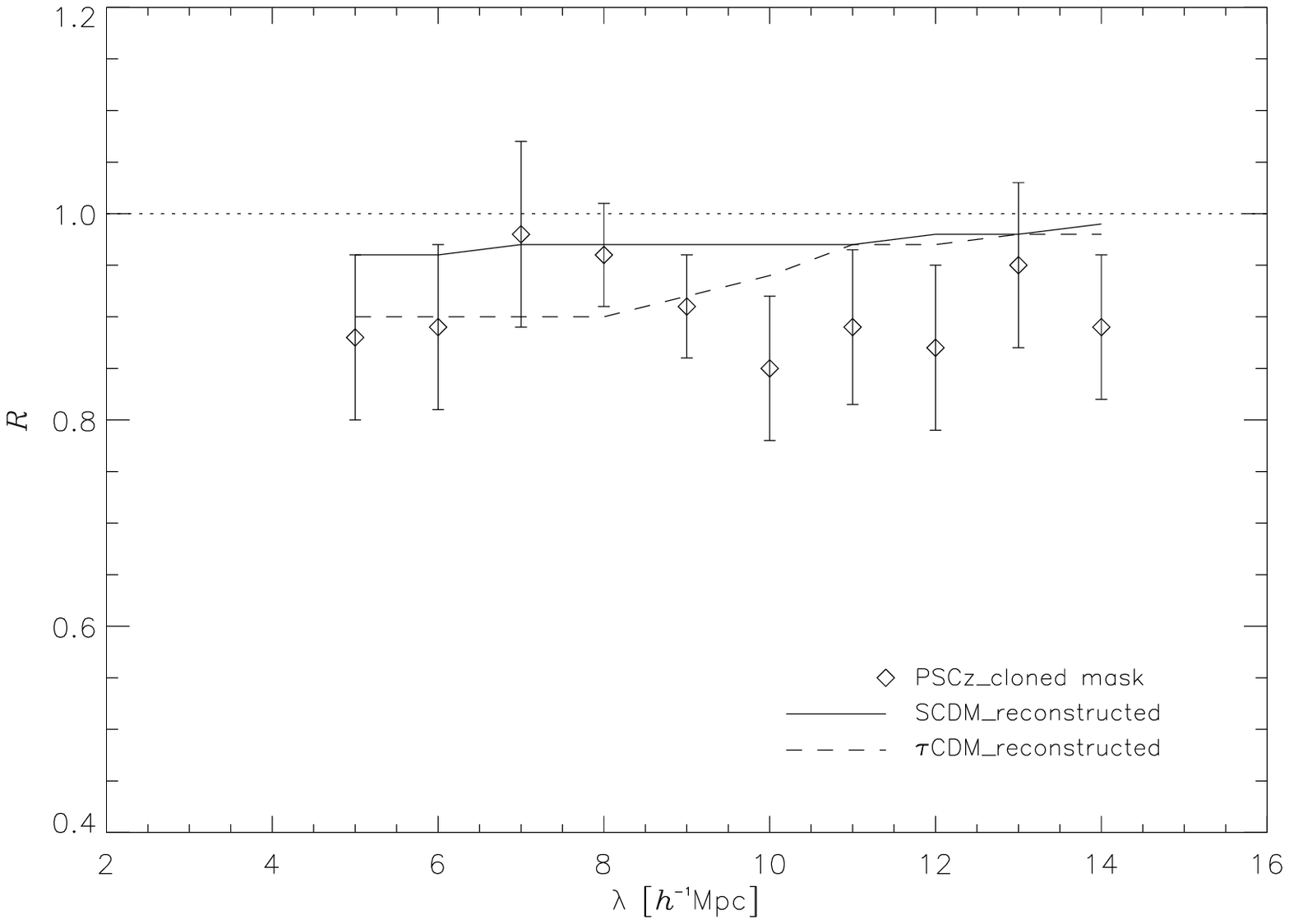}}
\resizebox{8cm}{!}{\includegraphics{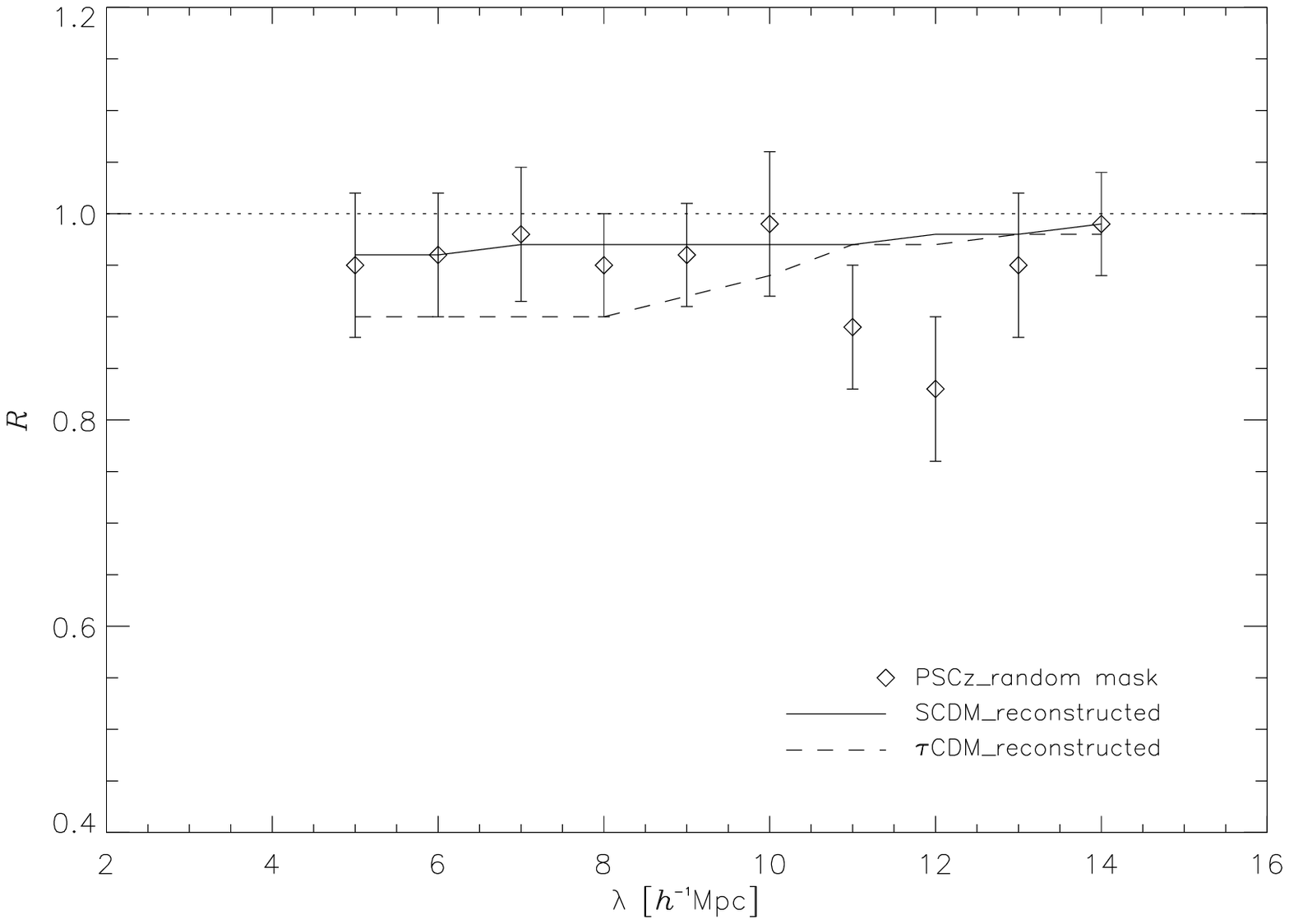}}
\caption{The amplitude drops measured for the reconstructed PSCz density fields. In the upper panel the mask has been cloned, whereas in the lower panel the mask has been filled randomly. Also shown are the amplitude drops obtained for the reconstructed N-body simulations. \label{psczdrops}}
\ec
\end{figure*}

\subsection*{Results \& Discussion} 

Fig. \ref{PSCzgenus} shows the genus curves obtained for the reconstructed PSCz fields. On the first column we plot the genus curves of the fields for which the mask has been filled using a cloning technique, and on the second column we plot the genus curves of the fields where the mask has been filled randomly. The thick solid lines refer to the reconstructed PSCz fields {\em proper}, whereas the thin solid lines refer to the randomized versions of those fields,i.e., the fields obtained from the reconstructed density fields by randomizing phases in Fourier space subject to the reality constraint $\delta_{{\bf k}}=\delta_{-{\bf k}}^{\ast}$ and keeping the same power spectrum. The dotted and dashed lines are the best fitting random-phase curves to both the reconstructed PSCz density fields and their randomized versions.

It is striking to notice the proximity between genus curves, even at small smoothing lengths. The statistical errors drawn from the N-body simulation are not shown in Fig. \ref{PSCzgenus} because of the high degree of correlation between the points in the genus curves.

Independently of whether the mask has been randomly filled or cloned, the recovered genus curves seem consistent with random-phase Gaussian fluctuations because the amplitude drops appear small. However, the amplitude themselves seem to depend on the mask filling technique.

In Fig. \ref{psczdrops} we plot the amplitude drops obtained for the reconstructed PSCz density fields, as a function of smoothing length, as well as the error bars drawn from the $\tau$CDM model following the method outlined in chapter \ref{recdenmaps}. The amplitude drops of the reconstructed N-body simulations are also shown for comparison.

In contrast with Fig. \ref{drpscznow}, where phase correlations were found for PSCz at small smoothing lengths, the reconstructed PSCz fields do not show any significant phase correlations on scales ranging from $5\lu$ to $14\lu$. This reinforces the hypothesis that density fluctuation originate from random-phase Gaussian initial conditions.

\section*{AKNOWLEDGEMENTS}

We are grateful to the Virgo consortium (J. Colberg, H. Couchman, G. Efstathiou, C. S. Frenk, A. Jenkins, A. Nelson, J. Peacock, F. Pearce, P. Thomas and S. D. M. White) for provinding simulation data ahead of publication. AC aknowledges the support of FCT (Portugal).

\bibliography{zeld}

\end{document}